\documentclass[11pt,a4paper]{article}
\pdfoutput=1

\usepackage[utf8]{inputenc} 
\usepackage{bbold}
\usepackage{slashed}
\usepackage{a4wide}
\usepackage{amsmath}
\usepackage{cite}       
\usepackage{xcolor}
\usepackage{amssymb}
\usepackage{amsfonts}
\usepackage{booktabs}
\usepackage{makecell}
\usepackage{multirow}
\usepackage{pdflscape}
\usepackage{textcomp}
\usepackage{gensymb}
\usepackage{bigstrut}
\usepackage{array}
\usepackage{enumerate}
\usepackage{enumitem}
\usepackage[small,bf]{caption}
\usepackage{subcaption}
\usepackage{diagbox}
\usepackage{graphicx}
\usepackage{mathbbol}
\usepackage{appendix}
\usepackage{verbatim}
\usepackage[top=2.5cm,bottom=2.5cm,left=3cm,right=3cm]{geometry}
\usepackage{makecell}
\usepackage{listings}
\usepackage{mathtools}
\usepackage{dsfont}
\usepackage{braket}
\usepackage{longtable}
\usepackage{threeparttable}
\usepackage{floatpag}
\usepackage[colorlinks=true,urlcolor=myred,anchorcolor=black,citecolor=myred,filecolor=black,linkcolor=myred,menucolor=blue,linktocpage=true]{hyperref}
\usepackage[normalem]{ulem}
\usepackage{multicol}

\usepackage[capitalise]{cleveref} 
\crefname{table}{Table}{tables}

\definecolor{myred}{rgb}{0.8, 0, 0.4}
\definecolor{myblue}{cmyk}{0.8, 0.4, 0, 0.2}
\definecolor{mygreen}{rgb}{0.27, 0.64, 0.28}
\definecolor{mygray}{gray}{.95}
\definecolor{verdes}{cmyk}{0.92,0,0.59,0.4}

\newcommand{\virg}[1]{``#1''}

\DeclareMathOperator{\diag}{diag}

\DeclareMathOperator{\im}{Im}
\DeclareMathOperator{\re}{Re}

\newcommand{\id}{\mathds{1}}

\def\Mav{M_\text{av}}

\newcolumntype{x}[1]{>{\centering\arraybackslash\hspace{0pt}}p{#1}}



\begin{document}

\vspace*{-15mm}
\begin{flushright}
\end{flushright}
\vspace*{5mm}

\vspace{2cm}

\renewcommand*{\thefootnote}{\fnsymbol{footnote}}

\begin{center}
{\bf 
{\LARGE Modular-symmetry-protected seesaw}
}\\[8mm]
A.~Granelli$^{\,a,b~}$\footnote{E-mail: \href{mailto:alessandro.granelli@unibo.it}{\texttt{alessandro.granelli@unibo.it}}},
D.~Meloni$^{\,c,d~}$\footnote{E-mail: \href{mailto:davide.meloni@uniroma3.it}{\texttt{davide.meloni@uniroma3.it}}},
M.~Parriciatu$^{\,c,d~}$\footnote{E-mail: \href{mailto:matteo.parriciatu@uniroma3.it}{\texttt{matteo.parriciatu@uniroma3.it}}},\\[1mm]
J.~T.~Penedo$^{\,d~}$\footnote{E-mail: \href{mailto:jpenedo@roma3.infn.it}{\texttt{jpenedo@roma3.infn.it}}} and
S.~T.~Petcov$^{\,e,f~}$\footnote{E-mail: \href{petcov@sissa.it}{\texttt{petcov@sissa.it}}. Also at Institute of Nuclear Research and Nuclear Energy, Bulgarian
Academy of Sciences, 1784 Sofia, Bulgaria.} \\
 \vspace{5mm}
$^{a}$\,{\it \small Dipartimento di Fisica e Astronomia, Università di Bologna,\\ Via Irnerio 46, 40126, Bologna, Italy} \\[1mm]
$^{b}$\,{\it \small INFN Sezione di Bologna, Viale Berti Pichat 6/2, 40127, Bologna, Italy} \\[1mm]
$^{c}$\,{\it \small Dipartimento di Matematica e Fisica, Università di Roma Tre,\\ Via della Vasca Navale 84, 00146, Roma, Italy} \\[1mm]
$^{d}$\,{\it \small INFN Sezione di Roma Tre, Via della Vasca Navale 84, 00146, Roma, Italy} \\[1mm]
$^{e}$\,{\it \small SISSA/INFN, Via Bonomea 265, 34136 Trieste, Italy} \\[1mm]
$^{f}$\,{\it \small Kavli IPMU (WPI), UTIAS, The University of Tokyo, Kashiwa, Chiba 277-8583, Japan} \\
\end{center}

\vskip 7mm      
\begin{center}
\textbf{Abstract}
\end{center}
\vspace{-.5em} 
In the presence of a finite modular flavour symmetry, fermion mass hierarchies may be generated by a slight deviation of the modulus from a symmetric point. We point out that this small parameter governing charged-lepton mass hierarchies may also be responsible for the breaking of lepton number in a symmetry-protected low-scale seesaw, sourcing active neutrino masses and the mass splitting of a pseudo-Dirac pair of heavy neutrinos.
We discuss the phenomenological implications of this mechanism, including the possibility to test the considered models at future planned and proposed heavy neutral lepton searches.

\vspace{1em}

\renewcommand*{\thefootnote}{\arabic{footnote}}
\setcounter{footnote}{0}

%%%%%%%%%%%%%%%%%%%%%%%
%
\clearpage
\vfill
\tableofcontents
\vskip 1cm
\hrule
\vskip 1cm
%%%%%%%%%%%%%%%%%%%%%%%%%%%%%%%%%%%%%%%%%%%%%%
\section{Introduction}
\label{sec:intro}
%%%%%%%%%%%%%%%%%%%%%%%%%%%%%%%%%%%%%%%%%%%%%%

There is still an interesting set of puzzles in modern particle physics. Among them, the origin of neutrino masses, their peculiar mixing pattern (as determined in neutrino oscillation experiments), and the remarkably pronounced mass hierarchies of charged leptons are striking examples. As it is widely known, these puzzles call for a beyond-the-Standard-Model explanation, and several approaches have been attempted in the past. It can be argued that the ultimate goal, at least from a bottom-up perspective, should be to solve all of them through a single mechanism. 
In doing so, one should also strive for minimality, predictivity and testability.

Regarding the generation of neutrino masses, one may focus on the simplest and most minimal scenario: the type-I seesaw~\cite{Minkowski:1977sc,Yanagida:1979as,Gell-Mann:1979vob,Glashow:1979nm,Mohapatra:1979ia}.
The type-I seesaw mechanism explains the smallness of neutrino masses thanks to right-handed neutrinos (RH$\nu$s), which are singlets under the Standard Model (SM) gauge group, possess Majorana masses and couple to left-handed lepton doublets via Yukawa interactions with the Higgs field. After electroweak symmetry breaking, the interplay between Dirac and Majorana neutrino mass terms generates light Majorana masses for the active flavour neutrinos, naturally suppressed by the scale of the RH$\nu$s, and new heavy (i.e.~much heavier than the eV scale) Majorana neutrinos.
In some of the original seesaw proposals based on Grand Unification
Theory (GUT) the heavy Majorana neutrino mass scale is close to the GUT
scale.
One motivation for the GUT-scale seesaw is the ``numerology argument''~\cite{altarelliquoteupdated} 
that provides the light neutrino mass scale of $m_\nu \sim 10^{-1\div 2}\,\text{eV}$ for $M\sim 10^{14\div15}\,\text{GeV}$. 
This may seem the most natural option, since in this way the Yukawa couplings (or Wilson coefficients) are $\mathcal{O}(1)$, but it renders the seesaw mechanism very
difficult (if not impossible) to test. 
On the other hand, the realization of the seesaw mechanism at sub-TeV scales is an appealing way of understanding the nature of neutrino masses, while also providing the opportunity to test it at, e.g., colliders and extracted beam-line facilities. In that context, the new heavy Majorana neutrinos that appear as a consequence of the seesaw mechanism constitute a class of Heavy Neutral Leptons (HNLs)  which can be produced directly in particle collisions, as well as through vector boson, tau lepton or  meson decays (see e.g.~\cite{Abdullahi:2022jlv, Antel:2023hkf} and references therein). However, to account for the smallness of light neutrino masses, the simplest realization usually implies highly-suppressed Yukawa couplings for the Dirac neutrino mass terms, going against the original intent of the seesaw mechanism.
Instead, various types of symmetry-protected seesaws have been proposed, e.g.~the ``inverse''~\cite{Wyler:1982dd,Mohapatra:1986bd} and ``linear''~\cite{Malinsky:2005bi} seesaws, (see  also~\cite{Shaposhnikov:2006nn,Kersten:2007vk,Gavela:2009cd,Ibarra:2010xw,Antusch:2015mia}) to make the sub-TeV scenario more natural. The feature of the symmetry-protected-seesaw is an additional symmetry, such as 
a ``lepton-number-like'' $U(1)_\text{L}$, under which both the RH$\nu$s and the lepton doublets are charged.
The small breaking of this symmetry participates in the suppression of light neutrino masses, thus alleviating the load on the suppression of the dimensionless Yukawa couplings.

Usually, the above
approach solves only one
piece of the puzzle, leaving no explanation for the specific pattern of neutrino mixing. One may argue that this other issue can be understood through another symmetry -- one regarding flavour. There have been several approaches based on flavour symmetries, both in the lepton and quark sectors (see~\cite{Ishimori:2010au,Altarelli:2010gt,King:2014nza,Petcov:2017ggy,Xing:2020ijf,Feruglio:2019ybq,Ding:2024ozt} for reviews).
In particular, modular flavour symmetries are a compelling candidate for an organizing principle~\cite{Feruglio:2017spp} (see~\cite{Kobayashi:2023zzc,Ding:2023htn} for recent reviews). In  this approach, the Yukawa couplings are modular forms, i.e.~holomorphic functions of the modulus $\tau$, a complex scalar spurion with $\im \tau >0$. Most entries of the mass matrices are then controlled by the vacuum expectation value (VEV) acquired by $\tau$, which breaks
modular invariance.
As a result, fermion mixing and other observables are reproduced up to a limited number of Lagrangian free parameters.
In minimal approaches, the VEV of the modulus can also be the only source of spontaneous CP violation~\cite{Novichkov:2019sqv}. 
On the other hand, modular flavour symmetries are considered an interesting framework also because they can be employed in understanding fermion mass hierarchies~\cite{Novichkov:2021evw}, 
originating as powers of a
small departure,
$|\epsilon| \ll 1$, of the VEV of $\tau$
from points of residual symmetries. 
For this reason, it can be argued that they represent one of the most promising frameworks to develop an elegant mechanism that relates all three aforementioned puzzles, which is also testable through the phenomenology of HNLs. Indeed, in this work we show how modular symmetry can act as the source of approximate lepton number conservation: a \emph{modular-symmetry-protected scenario where light neutrino masses and the HNL mass splittings} are suppressed by the same $\epsilon$ which governs charged-lepton mass hierarchies.

\vskip 2mm
We start by presenting the framework in~\cref{sec:framework}, reviewing the general symmetry-protected low-scale seesaw, the role of finite modular symmetries as flavour symmetries and how they can be responsible for fermion mass hierarchies. We then discuss how a modular-symmetry-protected setup can be realized. In~\cref{sec:models}, we consider specific models of this type and show under which conditions they agree with existing data. Then, in~\cref{sec:pheno}, we examine 
the HNL phenomenology of these models.
Finally, we summarize our results in~\cref{sec:summary}.

%%%%%%%%%%%%%%%%%%%%%%%%%%%%%%%%%%%%%%%%%%%%%%
\section{Framework}
\label{sec:framework}
%%%%%%%%%%%%%%%%%%%%%%%%%%%%%%%%%%%%%%%%%%%%%%

%%%%%%%%%%%%%%%%%%%%%%%%%%%%%%%%%%%%%%%%%%%%%%
	\subsection{Symmetry-protected low-scale seesaw}
	\label{sec:symprot}
%%%%%%%%%%%%%%%%%%%%%%%%%%%%%%%%%%%%%%%%%%%%%%

In order to generate light Majorana neutrino masses, we consider the type-I seesaw extension of the SM content by adding $n_R = 2$ or 3 RH$\nu$s, $\nu_{\kappa R}$, where $\kappa = 1, \ldots,n_R$. The RH$\nu$s are taken to be singlets under the SM gauge symmetry, entering the Lagrangian in general with a Majorana mass term and a Yukawa coupling to the Higgs and left-handed (LH) lepton doublets. At low energy, we then have:
\begin{equation}
\label{eq:Lseesaw}
{\cal L}\,\supset\, -(Y_e^*)_{\alpha \beta}\,\overline{\ell_{\alpha L}}\,\tilde{H}_d\,e_{\beta R} -
\,(Y_D^{*})_{\alpha \kappa} \,\overline{\ell_{\alpha L}}\,\tilde{H}_u\,\nu_{\kappa R}
 - \,\frac{1}{2}(M_R^*)_{\kappa\rho} \,\overline{\nu_{\kappa L}^c}\nu_{\rho R}+ \hbox{h.c.}\,,
\end{equation}
where we also show the charged-lepton Yukawa couplings, and we have anticipated the supersymmetric
origin of these terms. Here, the flavour-basis lepton doublet fields are $\ell_{\alpha L}^T \equiv (\nu_{\alpha L},\,e_{\alpha L}) $, while $\tilde{H}_{u,d}\equiv i\sigma_2 H_{u,d}^*$, where Higgs doublets take VEVs $\braket{H_{u}}=(0,\, v_u)^T$ and $\braket{H_{d}}=(v_d,\, 0)^T$. We recall that $\nu_{\kappa L}^c\equiv C (\overline{\nu_{\kappa R}})^T$, 
where $C$ is the charge conjugation matrix, and $M_R$ is symmetric. 
After electroweak symmetry breaking, the following neutrino mass terms are generated:
    \begin{equation}\label{eq:Lmnu}
-{\cal L}_{\nu} =\frac{1}{2} \,
    \overline{\mathcal{V}^c_R}\mathcal{M}\mathcal{V}_L+\text{h.c.}\equiv \frac{1}{2} \,
    \overline{\mathcal{V}^c_R}
\begin{pmatrix}
    \mathbb{0}&m_D\\[1mm]
    m_D^T&M_R
\end{pmatrix}
\mathcal{V}_L
+ {\rm h.c.}\,, 
\end{equation}
having defined the vector $\mathcal{V}_L^T \equiv (\nu_{\alpha L}, \nu_{\kappa L}^c)$.
Then, $m_D$ and $M_R$ denote the complex Dirac-type and Majorana-type neutrino mass matrices, respectively, with $(m_D)_{\alpha \kappa} = (Y_D)_{\alpha \kappa}\, v_u$.

\vskip 2mm

A symmetry-protected low-scale seesaw scenario (see e.g.~\cite{Antusch:2015mia}) relies on the small breaking of a lepton-number-like, or \virg{non-standard} lepton number~\cite{Leung:1983ti}, symmetry L.%
\footnote{The \virg{non-standard} 
lepton charge is expressed in terms of the individual lepton charges 
$L_\ell$ ($\ell=e,\mu,\tau$) and 
$L_a(\nu_{\kappa R}) \equiv -\,\delta_{a\kappa}$ ($a,\kappa=1,2$) as
$\text{L} = L_e + L_\mu + L_\tau - L_1 + L_2$. 
Then ${\rm min}(n_+,n_-)$ and $|n_+ - n_-|$
are the numbers of massive Dirac and massless neutrinos, respectively, 
$n_+$ ($n_-$) being the number of charges entering into the expression 
for L with positive (negative) sign~\cite{Leung:1983ti} (see also~\cite{Bilenky:1987ty}).
}
For definiteness, consider the assignments: $\text{L}(\nu_{1R}) = +1$, i.e.~$\nu_{1R}$ has the same charge as the SM LH doublets, while $\text{L}(\nu_{2R}) = -1$ and $\text{L}(\nu_{3R}) = 0$, so that $\nu_{3R}$ is decoupled from $\nu_{\alpha L}$ and $\nu_{1,2R}$.
Then, in the L-symmetric limit, the full neutrino mass matrix $\mathcal{M}$ becomes:
\begin{equation} \label{eq:6x6M}
\renewcommand*{\arraystretch}{1.2}
\mathcal{M}
\,\to\, 
\left(\begin{array}{ccc|ccc}
0 & 0 & 0 & y_{11} \, v_u & 0 & 0 \\
0 & 0 & 0 & y_{2 1} \, v_u & 0 & 0 \\
0 & 0 & 0 & y_{3 1} \, v_u & 0 & 0 \\
\hline
y_{11} \, v_u & y_{2 1} \, v_u & y_{3 1}\,  v_u & 0 & M_{12} & 0 \\
0& 0 & 0 & M_{12} & 0 & 0 \\
0& 0 & 0 & 0 & 0 & M_3
\end{array}\right)
\,.
\end{equation}
In this limit of a conserved L charge, the three active neutrinos are massless, the RH$\nu$s charged under L (namely $\nu_{1 R}$ and $\nu_{2R}$) make up a Dirac particle with mass $M_{12}$, while the $\nu_{3R}$ field corresponds to a decoupled Majorana neutrino with mass $M_3$.
In the minimal case with 2 RH$\nu$s, which is sufficient to explain neutrino oscillation data, the neutrino mass matrix $\mathcal{M}$ in the symmetric limit is simply given by~\cref{eq:6x6M} with the last row and the last column removed. We refer to this specific structure of the $2\times 2$ Majorana mass matrix as \emph{Pauli-like}.

A small breaking of the L symmetry, $\epsilon$, is required to generate active neutrino masses as well as mixing in the lepton sector. Its introduction generically lifts the mass degeneracy of the Dirac fermion built from $\nu_{1,2R}$, which instead becomes a pseudo-Dirac pair~\cite{Wolfenstein:1981kw,Petcov:1982ya}.
The sub-eV scale of active neutrino masses then emerges from the interplay of two effects: i) the small breaking $\epsilon$ of the L symmetry and ii) the traditional suppression through the heavy masses of the HNLs, which can however be in the GeV range. For concreteness, let
\begin{equation}
 m_D=m_0+\epsilon^d\, m_1\,,\qquad   
 M_R=M_0+\epsilon^r\, M_1\,,
\end{equation}
where $d$ and $r$ are the smallest exponents for the powers of $\epsilon$ in the perturbation matrices $m_1$ and $M_1$. 
Now, $m_0$ and $M_0$ are the Dirac and Majorana matrices in the symmetric limit ($\epsilon \to 0$) illustrated in~\eqref{eq:6x6M}. In the seesaw limit ${m_D}\ll {M_R}$ (at the level of singular values), block diagonalization of the mass terms in~\cref{eq:Lmnu} leads to the type-I seesaw formula for the mass matrix of light neutrinos:%
\footnote{In our convention, the active flavour neutrino mass term reads $\mathcal{L}\supset -\frac{1}{2}(m_\nu)_{\alpha \beta}\overline{\nu_{\alpha R}^c}\nu_{\beta L}+\text{h.c.}$.}
\begin{equation} \label{eq:seesawtypeI}
\begin{aligned}
  m_\nu &\,\simeq\, -m_D\, M_R^{-1}\, m_D^T\\[1mm]
   &\,=\,
     \epsilon^r\, m_0 M_0^{-1}M_1 M_0^{-1}m_0^T
    -(\epsilon^d\, m_1 M_0^{-1}m_0^T
    +\text{transpose})\,+\,\mathcal{O}\big(|\epsilon|^{2r},|\epsilon|^{2d},|\epsilon|^{r+d}\big)
    \,,
\end{aligned}
\end{equation}
where we used the fact that $m_0 M_0^{-1} m_0^T$ is the zero matrix. As it can be seen, the same perturbation $\epsilon^r M_1$, which is predominantly responsible for the breaking of the mass degeneracy of the $\nu_{1R}$ and $\nu_{2R}$ states, also contributes to the lifting of the light neutrino masses from zero. In terms of orders of magnitude, the active neutrino mass scale follows, at leading order, from:
\begin{equation} \label{eq:numass}
    m_{i}^\nu \,\sim\, \max(|\epsilon|^r,|\epsilon|^d)\times\frac{y^2v_u^2}{M}\,,
\end{equation}
where $y^2$ is some real combination of Yukawa couplings (or Wilson coefficients), and $M$ is a mass scale associated with $M_R$. As a result, the sub-eV scale of light neutrinos can be achieved with a less severe suppression of the dimensionless parameters of the theory.\footnote{\label{foot:loop}%
In order to reach the sub-eV active neutrino mass scale, the overall size of the Yukawas must still be suppressed (in our models we find $y< \mathcal{O}(10^{-4})$), and
a valid seesaw expansion is still guaranteed. Thus, our conclusions are not spoiled by next-to-leading seesaw terms in an expansion in $y\, v_u / M$ (see e.g.~\cite{Dubinin:2023yli}). Our results should also be stable with respect to loop corrections (both supersymmetric and standard contributions), which generally scale as $\sim y^4 v_u^2/M$ multiplied by the usual loop factors, see e.g.~\cite{SuarezNavarro:2024wlj}. For $M\in [10^{-1},10^2]$ GeV as considered in this work, such stability can be achieved by requiring $y\lesssim 10^{-4}$, which is always satisfied in our models, as anticipated.}
It follows from~\cref{eq:numass} that,
given $m^\nu_i$ and $M$,
the neutrino Yukawa coupling $y$
is enhanced with respect to the one in the standard seesaw scenario by the
factor $\max(|\epsilon|^r,|\epsilon|^d)^{-1/2}$.

The diagonalization of the full mass matrix in~\cref{eq:Lmnu} also gives three additional states, corresponding to Majorana neutrinos that are heavier than the active ones. The mass matrix of these heavy Majorana neutrinos, $M_N$, can be written as $M_R$ plus 
corrections (see e.g.~\cite{Penedo:2017knr}):
\begin{equation} \label{eq:Mmass}
\begin{aligned}
M_N&\simeq M_R+\frac{1}{2}\left((M_R^*)^{-1}m_D^\dagger m_D+\text{transpose}\right) \\
&=M_0+\epsilon^r\, M_1+\frac{1}{2}\left(\delta M_0+\delta M_0^T\right)\\
&\quad +\frac{1}{2}\left(\epsilon^{d*}\,\delta M_a+\epsilon^{d}\,\delta M_b-\epsilon^{r*}\,\delta M_c+\text{transpose}\right)+\mathcal{O}\big(|\epsilon|^{2r},|\epsilon|^{2d},|\epsilon|^{r+d}\big)\,,
\end{aligned}
\end{equation}
where $\delta M_0\equiv (M_0^*)^{-1} m_0^\dagger m_0$ is the contribution from the Dirac mass term in the symmetric limit, which, however, does not provide the splitting $\Delta M$ of $\nu_{1R}$ and $\nu_{2R}$, as can be easily verified. We have also defined: 
\begin{align}
\delta M_a    &\equiv (M_0^*)^{-1}m_1^\dagger m_0\,,\\
\delta M_b &\equiv (M_0^*)^{-1}m_0^\dagger m_1\,,\\
\delta M_c &\equiv (M_0^*)^{-1}M_1^*(M_0^*)^{-1} m_0^\dagger m_0
    \,.
\end{align}
From~\eqref{eq:Mmass} we can infer the mechanism behind the splitting of the heavy Dirac pair. 
If $M_1$ is non-zero, it is expected to provide the leading contribution to the splitting,
\begin{equation}    \label{eq:heavysplitting0}
    \Delta M\sim |\epsilon|^r\times M\,.
\end{equation}
If, instead,
the symmetry is not explicitly broken in the heavy Majorana sector, namely if $M_1$ is the zero matrix, the splitting at leading order follows from:
\begin{equation}  \label{eq:heavysplitting1}
    \Delta M\sim |\epsilon|^d\times \frac{y^2 v_u^2}{M}\,,
\end{equation}
thus, we obtain a splitting proportional to the size of the light neutrino mass scale (note that $\delta M_c = \mathbb{0}$ in this case).

\vskip 2mm
In what follows, we will show how residual modular symmetries may play the role of the L symmetry. In such a case, their small breaking not only underlies the origin of neutrino masses and mixing -- it can also be responsible for the observed hierarchies between the masses of charged leptons, whose spectrum is governed by powers of the same $\epsilon$.

%%%%%%%%%%%%%%%%%%%%%%%%%%%%%%%%%%%%%%%%%%%%%%
	\subsection{Modular flavour symmetries}
	\label{sec:modframework}
%%%%%%%%%%%%%%%%%%%%%%%%%%%%%%%%%%%%%%%%%%%%%%

In the minimal realization of modular symmetry, the VEV of the modulus $\tau$ can act as the single source of flavour symmetry breaking and CP violation.
In part, the predictive power of the modular setup relies on holomorphicity, which calls for a supersymmetric (SUSY) framework. Thus, modular symmetry can significantly constrain the superpotential $W$ of the theory.\footnote{It must be noted that, on the other hand, the Kähler potential is not constrained by modular symmetry~\cite{Chen:2019ewa},  
because it is non-holomorphic. In the absence of a top-down prescription, one can assume a minimal expression for the Kähler potential, which we show in~\cref{eq:kahler}. However, it has been argued that the eventual corrections, in some approaches, can be made comparable to the experimental uncertainties of lepton observables~\cite{Chen:2021prl}.} 
In practice, low-energy predictions can remain unaffected by the dynamics of SUSY breaking in regions of the parameter space where a sufficient separation between the scales of soft terms and of SUSY-breaking mediation can be arranged~\cite{Criado:2018thu}.

Under a modular transformation $\gamma \in \Gamma$, where $\Gamma = SL(2,\mathbb{Z})$ is the modular group, the modulus transforms according to 
\begin{equation}
\tau \to \frac{a\tau +b}{c\tau+d}\,, \qquad \text{with }\gamma = \bigg(\begin{array}{cc} a & b \\ c & d\end{array}\bigg)\,,
\end{equation}
where $a,b,c,d \in \mathbb{Z}$ satisfying $ad-cb=1$.
The modular group is generated by the transformations
 $S: \tau \to -1/\tau$ and $T: \tau \to \tau +1$, and can be presented via the relations $(ST)^3 = S^4 =\id$ and $S^2T = TS^2$. A modular form is a holomorphic function of $\tau$ which under $\Gamma$ transforms as:
 \begin{equation}
     \label{eq:mformtr}
     f(\tau)\to (c\tau+d)^kf(\tau)\,,
 \end{equation}
 where $k$ is a positive integer called ``weight''. In this context, superfields $\psi$ transform as 
\begin{equation}
\psi \to (c\tau + d)^{-k_\psi} \rho_\mathbf{r}(\gamma) \psi
\end{equation}
under the modular flavour symmetry. Note that, despite being commonly called ``weight'', the integer $k_\psi$ for the matter superfields has no sign constraints. That is, matter superfields are not modular forms. On the other hand, the (unitary) representation matrix $\rho_\mathbf{r}$ is a representation of a finite non-Abelian group $\Gamma_N' = \Gamma/\Gamma(N)$, with the natural number $N=2,3,\ldots$ being its so-called level.
Here, $\Gamma(N)$ is the principal congruence subgroup of $SL(2,\mathbb{Z})$ of matrices equal to the unit $2\times 2$ matrix modulo $N$.
For small values of $N$, these finite modular groups are isomorphic to the traditional flavour symmetry groups or to their ``double covers'': $\Gamma_2' \simeq S_3$, $\Gamma_3' \simeq A_4' \equiv T'$, $\Gamma_4' \simeq S_4'$, and $\Gamma_5' \simeq A_5'$.
To be able to write a modular-invariant superpotential, one makes use of modular forms $Y(\tau) \equiv Y_{\mathbf{r}_Y}^{(k_Y)}(\tau)$, which, under $\Gamma_N'$, transform as
\begin{equation}
    Y(\tau) \to (c\tau + d)^{k_Y} \rho_{\mathbf{r}_Y}(\gamma)\, Y(\tau)\,.
\end{equation}
For a fixed weight $k_Y$, the set of non-zero modular forms is finite, limiting the possible terms one can add to $W$.

Any VEV of $\tau$ breaks the full modular group $\Gamma$. Usually, in the bottom-up approach we are following, it is enough to scan $\tau$ in the fundamental domain $\mathcal{D}$ in order to fit a minimal (no-flavons) modular-invariant model to data, from the bottom-up. Said region corresponds to $|\tau| \geq 1$ and $-1/2 \leq \re\tau \leq 1/2$.
One may also impose a generalized CP (gCP) symmetry on the theory. Then, the VEV of $\tau$ can also be the only source of CP violation (CPV). Under gCP, the modulus transforms as $\tau \to - \overline\tau$~\cite{Baur:2019kwi,Novichkov:2019sqv}. Moreover, there is a basis in which $\psi \to \overline\psi$ and $Y(\tau) \to Y^*(\tau)$ under gCP~\cite{Novichkov:2019sqv}. As a result, all the remaining free parameters in the superpotential must be real.
On the other hand, there exist special values of $\tau$ that preserve the CP symmetry. Inside $\mathcal{D}$, these are $|\tau| = 1$ and $|\!\re\tau| = \{0, \,  1/2\}$.

%%%%%%%%%%%%%%%%%%%%%%%%%%%%%%%%%%%%%%%%%%%%%%
	\subsubsection*{Hierarchies from residual modular symmetries}
	\label{sec:hierarchies}
%%%%%%%%%%%%%%%%%%%%%%%%%%%%%%%%%%%%%%%%%%%%%%

At three specific values (fixed points) of the VEV of 
$\tau \in \mathcal{D}$, the modular symmetry groups  
$\Gamma = SL(2,\mathbb{Z})$ and  $\Gamma_N'$ 
are  broken only partially 
to Abelian discrete residual symmetries:
i) at  $\tau  = e^{2\pi i/3} \equiv \omega$ 
to $\mathbb{Z}^{ST}_3$, 
ii) at $\tau = i\infty$ to  $\mathbb{Z}^{T}_N$, and
iii) at  $\tau = i$ to $\mathbb{Z}^{S}_2$. 
Some modular forms vanish  
at these symmetric points. As a consequence, 
several entries in the
fermion mass 
matrices may also vanish. 
In the \virg{proximity} of $\tau$ to a given 
symmetric point (e.g.~$\tau \simeq \omega$), 
parameterized by a small \virg{distance}  $\epsilon$, 
these zeros are lifted by powers $\epsilon^n$, which can lead to hierarchical 
fermion mass spectra~\cite{Novichkov:2021evw}.%
\footnote{For each entry of the mass matrix, the exponents $n$ are determined by the group $\Gamma_N'$, the field irreducible representation and weight assignments $(\mathbf{r},k)$, and the residual symmetry at the symmetric point.}
In particular, a non-fine-tuned fit of neutrino data with charged-lepton mass 
hierarchies originating from this mechanism has been found in 
Ref.~\cite{Novichkov:2021evw} for $\tau \simeq \omega$.
While stabilizing $\tau$ at a small but non-zero distance from a symmetric 
point may seem {\it ad hoc}, 
simple modular-invariant potentials for $\tau$ can
have non-fine-tuned minima in the requisite close 
vicinity of $\omega$~\cite{Novichkov:2022wvg}.
In minimal models without flavons,
the hierarchical charged-lepton (or quark) 
mass spectrum can naturally be obtained without fine-tuned
parameters
also for the VEV of $\tau$ in the \virg{vicinity} 
of the fixed point  $\tau=i\infty$~\cite{Novichkov:2021evw}.

%%%%%%%%%%%%%%%%%%%%%%%%%%%%%%%%%%%%%%%%%%%%%%%%%%%%%%%%%%%%%%%%%%%%%%%%%%%%%%%%%%
\subsection{Modular symmetry-protected low-scale seesaw}
\label{sec:modsymprot}
Denoting the superfields associated with the 
RH neutrinos
by $N^c$, the lepton doublet superfields by $L$ and the charged-lepton singlet superfields by $E^c$, the lepton-sector superpotential reads:
%%%%%%%%%%%%%%%%%%
\begin{equation} 
\label{eq:W}
W =  (Y_e)_{\alpha\beta} L_\alpha E^c_\beta H_d + (Y_D)_{\alpha\kappa} L_\alpha N^c_\kappa H_u+ \frac{1}{2} (M_R)_{\kappa\rho} N^c_\kappa N^c_\rho \,.
\end{equation}
%%%%%%%%%%%%%%%%%%%%%%%%%%
%
At low energies it leads  to the Lagrangian terms of~\cref{eq:Lseesaw}. In the present study we assume the minimal form for the Kähler potential:
%%%%%%%%%%%%%%%%%%%%%%
\begin{equation}
\label{eq:kahler}
K(\Phi,\bar\Phi)=-h\Lambda_\tau^2 \log(-i\tau+i\bar\tau)+\textstyle\sum_I(-i\tau+i\bar\tau)^{-k_I}{|\varphi_{I}|^2}\,,
\end{equation}
%%%%%%%%%%%%%%%%%%%%%%%%%%
%
which gives rise to the kinetic terms. Here, $h>0$ and $\Lambda_\tau$ has mass-dimension of one, and we denote with $\varphi_{I}$ all the matter superfields.
After the modulus $\tau$ acquires a VEV,
these fields will need to be rescaled as $\varphi_I  \to (2\,\im\tau)^{k_I/2}\, \varphi_I$ to yield canonical kinetic terms.
Hence, the original superpotential parameters, which hereafter are denoted with hats and in~\cref{eq:W} are included in $Y_e$ and $Y_D$, are rescaled by the relevant powers of $2\,\im\tau$. In~\cref{sec:modelsubsec}, we drop hats to indicate that this rescaling has already taken place (see the expressions for $Y_e$ and $Y_D$ in that section).
In the following, we will also neglect the dynamics associated with the superpartners, since we assume a high SUSY-breaking scale. As shown in~\cite{Criado:2018thu}, SUSY-breaking corrections to predictions obtained in modular-symmetric models can be made negligible under fairly mild assumptions.

\vskip 2mm

As it was previously discussed in~\cref{sec:symprot}, one wants to depart from the symmetric configuration of~\eqref{eq:6x6M} by means of a small quantity $\epsilon$, and in general the mass matrices $m_0$ and $M_0$ will receive corrections as $m_0+\epsilon^d\, m_1$ and $M_0+\epsilon^r\, M_1$. 
We now go beyond the existing literature by proposing the modular version of the symmetry-protected seesaw mechanism, which features the following aspects:
\begin{enumerate}[label=(\alph*)]
    \item The structures of the matrices $m_1$ and $M_1$ are determined by the particular arrangement of the components of the modular forms. This is achieved through a judicious choice of the modular group and of the irreducible representations of the superfields. 
    \item The modular symmetry mimics the L symmetry only in the symmetric or fixed points $\tau_\text{sym} \in \{\omega,i,i\infty\}$. In any other point of the domain $\mathcal{D}$, the L symmetry is explicitly broken. The (assumed small) breaking of such symmetry is encoded in the “distance" to the fixed points, parameterized by $\epsilon$, which is $\tau$-dependent.
    \item The quantity $\epsilon$ is the same one that simultaneously provides for the charged-lepton mass hierarchies through the proximity to the fixed points, as described in the previous section.
\end{enumerate}

To realize this program, one is free to choose among the different finite modular groups $\Gamma_N'$ and their eventual extensions. 
In the absence of an organizing principle from top-down approaches, we make use of the following phenomenology-inspired decision process.

\begin{enumerate}[label=\textbf{Step} \arabic*, start=0]
    \item The group $\Gamma_N'$ must provide for a reliable realization of the charged-lepton mass hierarchies in the vicinity of the symmetric points. Since in our approach everything is controlled by the modular forms, we argue that the hierarchical structure of the mass matrices can be considered reliable only when it is not mainly dictated by the overall normalizations of the modular forms (or by the magnitudes of the superpotential parameters), but by the relative sizes of the different components inside the same multiplet.\footnote{This issue has been thoroughly discussed in section 2.2.1 of Ref.~\cite{deMedeirosVarzielas:2023crv}.} Adopting this viewpoint, one is led to exclude $\Gamma_N'$ groups that do not furnish \textbf{triplet} representations. At this level, such triplets can be assigned either to the $E^c$ isosinglets or to the $L$ isodoublets.
    \item The remaining groups from the previous step must now allow for the realization of a \textbf{Pauli-like} structure within
    the Majorana mass matrix $M_0$, in the limit of L symmetry, i.e., at the chosen fixed point. Following an analogous point of view to the one used for charged leptons in the previous step, this should be achieved only by assigning irreducible representations with dimensions greater than one to the superfields $N^c$. We argue that only in this way can we consider the Pauli-like structure to be reliably dictated by modular symmetry, and not by \textit{ad hoc} adjustments of free parameters. This step loosely indicates the assignments of modular weights and irreducible representations (irreps) for the $N^c$ superfields, depending on the available modular forms of the chosen group, and their behaviour near the fixed points.
    \item After the previous filter, the remaining groups must allow to realize the structure of the \textbf{Dirac} matrix $m_0$ in the limit of exact L symmetry.
    Like the previous steps, this is a non-trivial requirement, strongly dependent on the behaviour of the decompositions under the residual symmetry groups associated to each of the fixed points. For such reason, this fixes the representations of the electroweak left-handed doublets.
    \item At this stage, one needs to verify that the chosen representations and modular weights allow the construction of the 
    \textbf{charged-lepton} mass hierarchies through the residual symmetry approach.
    For example, one can have:
    \begin{equation} \label{eq:ex_Me}
M_e \sim \begin{pmatrix}
1 & 1 & 1 \\
\epsilon & \epsilon  & \epsilon  \\
\epsilon^2 & \epsilon^2 & \epsilon^2
\end{pmatrix}
\,,
\end{equation}
or its transpose, with $\epsilon \sim 10^{-2}$
in order to reproduce the observed hierarchies. Here, each column is a triplet of modular forms (or a combination thereof), and the indicated behaviour 
refers to the vicinity of the symmetric point, in an appropriate flavour basis. 
     
\end{enumerate}

In addition, since the number of modular forms of weight $k_Y$ grows linearly with $k_Y$, in order to retain only a limited number of free parameters in the superpotential, one may choose to limit the weights. This is done to further increase the predictive power of the models.
In practice, we bound the weights by the smallest possible value which still yields models ($k_Y \leq 5$ or 6).
Depending on the specific modular forms available at each weight, and for each $\Gamma_N'$, this requirement places a potentially strong constraint on the possible symmetry-protected models, due to the quite strict filters in the previous steps.  Regarding the breadth of our investigation, we choose to explore the first three $\Gamma_N'$ groups that furnish triplet representations, i.e.~$\Gamma_3'\simeq T'$, $\Gamma_4'\simeq S_4'$ and $\Gamma_5'\simeq A_5'$.

Finally, we note that, in the presence of reducible representations among different families, each matrix column or row may potentially emerge from different modular form multiplets, for whose absolute normalization there is no top-down prescription. In order to safeguard the reliability of the obtained matrix structures, one should employ a definite prescription to control the relative sizes of these columns or rows. To this end, we choose to normalize the modular forms according to Ref.~\cite{Petcov:2023fwh}, see~\cref{app:forms}. 

\vskip 2mm
 
In~\cref{tab:modelsA4,tab:modelsS4} of~\cref{app:models}, we list all potential models passing our decision process, as well as the expected magnitude of the HNL splitting $\Delta M$ within each model. We exclude from our list those models where at least one of the three matrices $Y_e$, $Y_D$, or $M_R$ vanishes in the L-symmetric limit.
The surviving models are either i) based on $\Gamma_3'\simeq T'$, with $\tau \simeq \{\omega, i\infty\}$ and leading to $\Delta M \sim  \epsilon\, M$ or $\Delta M\sim m_\nu$ (cf.~\cref{tab:modelsA4}), or ii) based on $\Gamma_4'\simeq S_4'$, with $\tau \simeq i\infty$ which provides $\Delta M \sim  \epsilon^2\, M$ or $\Delta M\sim m_\nu$ (cf.~\cref{tab:modelsS4}).
No $A_5'$-based models are permitted, irrespective of the upper bound on $k_Y$. 
Note that, even if a model is included in these tables, a fit to lepton data is not guaranteed.
In what follows, we focus on four benchmark models that successfully accommodate charged-lepton masses and oscillation data, and explore their phenomenological predictions. 

%%%%%%%%%%%%%%%%%%%%%%%%%%%%%%%%%%%%%%%%%%%%%%
\section{Benchmarks vs.~data}
\label{sec:models}

%%%%%%%%%%%%%%%%%%%%%%
\begin{table}[t]
\centering
\renewcommand{\arraystretch}{1.5}
\begin{tabular}{ccccccc} 
\toprule
Model & Group & $\tau_\text{sym}$ & $L$ & $E^c$ & $N^c$ & $\Delta M$
 \\ 
\midrule
 A & \multirow{2}{*}{$A_4$} & \multirow{2}{*}{$\omega$}  & $(\mathbf{1}, +2) \oplus(\mathbf{1'}, +4) \oplus(\mathbf{1''}, +6)$ &
 $(\mathbf{3},0)$ & $(\mathbf{3},0)$ &  $\sim m_\nu$
 \\
 B &  & & $(\mathbf{1'}, +1) \oplus(\mathbf{1''}, +3) \oplus(\mathbf{1}, +5)$ & $(\mathbf{3},+1)$ & $(\mathbf{3},+1)$ &  
 $\sim \epsilon\, M$
 \\
\midrule
 C & \multirow{2}{*}{$S_4'$} &  \multirow{2}{*}{$i\infty$} &$(\mathbf{\hat{1}}, +2) \oplus(\mathbf{1}, +3) \oplus(\mathbf{\hat{1}}, +4)$ & \multirow{2}{*}{$(\mathbf{3},+1)$} & \multirow{2}{*}{$(\mathbf{3},+1)$}  & \multirow{2}{*}{$\sim \epsilon^2\, M$}
  \\
 D & &  & $(\mathbf{\hat{1}'}, +2) \oplus(\mathbf{1}, +3) \oplus(\mathbf{\hat{1}'}, +4)$  & &   & 
  \\
\bottomrule
\end{tabular}
\caption{
Summary of benchmark models. For each model, we specify the finite modular group, the value of the symmetric point, with $\tau \simeq \tau_\text{sym}$, the modular assignments of lepton superfields, with $\psi \sim (\mathbf{r},k_\psi)$, and the expected magnitude of the HNL splitting $\Delta M$.
}
\label{tab:benchmarks}
\end{table}
\renewcommand{\arraystretch}{1.0}
%%%%%%%%%%%%%%%%%%%%%%
%

\subsection{Models}
\label{sec:modelsubsec}
In this section, we analyse the benchmark models identified in~\cref{tab:benchmarks} and assess their compatibility with the charged-lepton mass ratios as well as oscillation data.
As indicated, models A and B
are based on the finite modular group $\Gamma_3\simeq A_4$ 
with values of the modulus $\tau \simeq \omega$
and include $n_R = 3$ RH$\nu$s. In both models, all three 1D irreps of $A_4$ are used. Note that, for these particular benchmarks, it is sufficient to consider the modular forms and irreps of $\Gamma_3 \simeq A_4$, i.e.~it is not necessary to consider its double cover.
Models C and D also involve 3 RH$\nu$s, but both rely on the (double cover) finite modular group $\Gamma_4' \simeq S'_4$ and require a modulus with a relatively large imaginary part, $\tau \simeq i \infty$.\footnote{\label{foot:imtau}%
In the modular framework, since  $\tau$ appears inside complex exponentials, this usually amounts to asking $\text{Im}\,\tau \gtrsim 2$.}
In all four benchmark models, each generation of lepton isodoublets is assigned to a 1D irrep of the flavour group ($L = L_1 \oplus L_2\oplus L_3$), while $E^c$ and $N^c$ furnish 3D and $n_R$-D flavour irreps, respectively.

To increase the predictive power of these benchmark models, we impose the gCP symmetry on each one of them, so that all superpotential parameters are real (in an appropriate basis) and the VEV of $\tau$ sources both modular symmetry breaking and CP violation.
We thus make use of the symmetric bases for group generators and Clebsch-Gordan coefficients given in Refs.~\cite{Novichkov:2019sqv, Novichkov:2020eep}, for the $A_4$ and $S_4'$ benchmark models, respectively.

\subsubsection*{Model A}

In this model, the neutrino Majorana mass matrix  is $\tau$-independent and simply reads
%%%%%%%%%%%%%%%%%%%%%%%%
\begin{equation} 
\label{eq:majA40}
    M_R=\Lambda\begin{pmatrix}
   1 & 0 & 0 \\
   0 & 0 & 1 \\
   0 & 1 & 0
\end{pmatrix}\,,
\end{equation}
%%%%%%%%%%%%%%%%%%%
%
where $\Lambda$ has the dimension of a mass and corresponds to the HNL mass scale.
Clearly, the heavy neutrinos are degenerate in the limit of vanishing $Y_D$. In the weaker $\epsilon\to 0$ limit, two of the heavy Majorana states are still degenerate and combine to form a Dirac pair. For $\epsilon \neq 0$, the latter is split into a pseudo-Dirac pair with $\Delta M \sim \epsilon\, m_\nu$ (cf.~\cref{eq:limitYDA1} later).  

The neutrino Yukawa coupling matrix is given by
\begin{equation}
Y_D =  \begin{pmatrix}
g_1\left(Y^{(2)}_{\mathbf{3}}\right)_1 &
g_1\left(Y^{(2)}_{\mathbf{3}}\right)_3 &
g_1\left(Y^{(2)}_{\mathbf{3}}\right)_2\\[2mm]
g_2\left(Y^{(4)}_{\mathbf{3}}\right)_3 &
g_2\left(Y^{(4)}_{\mathbf{3}}\right)_2 &
g_2\left(Y^{(4)}_{\mathbf{3}}\right)_1 \\[2mm]
g_{3,1}\left(Y^{(6)}_{\mathbf{3},1}\right)_2 + g_{3,2}\left(Y^{(6)}_{\mathbf{3},2}\right)_2 &
g_{3,1}\left(Y^{(6)}_{\mathbf{3},1}\right)_1 + g_{3,2}\left(Y^{(6)}_{\mathbf{3},2}\right)_1 &
g_{3,1}\left(Y^{(6)}_{\mathbf{3},1}\right)_3 + g_{3,2}\left(Y^{(6)}_{\mathbf{3},2}\right)_3
\end{pmatrix}\,,
\end{equation}
where the $g_i$ are, in general complex, superpotential parameters.
Explicit $q$-expansions for the modular form multiplets, with $q\equiv e^{2\pi i \tau}$, are given in~\cref{app:forms}.
The charged-lepton Yukawa coupling matrix has exactly the same structure as $Y_D$, namely
\begin{equation} \label{eq:YeA}
Y_e =  \begin{pmatrix}
\alpha_1\left(Y^{(2)}_{\mathbf{3}}\right)_1 &
\alpha_1\left(Y^{(2)}_{\mathbf{3}}\right)_3 &
\alpha_1\left(Y^{(2)}_{\mathbf{3}}\right)_2\\[2mm]
\alpha_2\left(Y^{(4)}_{\mathbf{3}}\right)_3 &
\alpha_2\left(Y^{(4)}_{\mathbf{3}}\right)_2 &
\alpha_2\left(Y^{(4)}_{\mathbf{3}}\right)_1 \\[2mm]
\alpha_{3,1}\left(Y^{(6)}_{\mathbf{3},1}\right)_2 + \alpha_{3,2}\left(Y^{(6)}_{\mathbf{3},2}\right)_2 &
\alpha_{3,1}\left(Y^{(6)}_{\mathbf{3},1}\right)_1 + \alpha_{3,2}\left(Y^{(6)}_{\mathbf{3},2}\right)_1 &
\alpha_{3,1}\left(Y^{(6)}_{\mathbf{3},1}\right)_3 + \alpha_{3,2}\left(Y^{(6)}_{\mathbf{3},2}\right)_3
\end{pmatrix}\,,
\end{equation}
but depends on a different set of superpotential parameters, $\alpha_i$. We recall that both the $g_i$ and $\alpha_i$ parameters include the factors $(2\text{Im}\tau)^{k_Y/2}$, arising from the rescaling of the relevant fields. 

In an appropriate ($ST$-diagonal) flavour basis,%
\footnote{\label{foot:ST}%
To perform this change of basis, one needs to rotate the $A_4$ triplets, including the modular forms given in~\cref{eq:A4triplets} of~\cref{app:forms}, via $\mathbf{3} \to U_\mathbf{3} \,\mathbf{3}$, with $$U_\mathbf{3} = 
\frac{1}{3}\begin{pmatrix}
    -\omega^2 & 2 \omega & 2 \\
    2\omega^2 & -\omega & 2 \\
    -2\omega^2 & -2\omega & 1 
\end{pmatrix}\,.$$}
one has
\begin{equation} \label{eq:limitYDA1}
Y_D' \sim \begin{pmatrix}
\epsilon^2 & 1 & \epsilon \\
\epsilon^2 & 1 & \epsilon \\
\epsilon^2 & 1 & \epsilon 
\end{pmatrix}\,,
\end{equation}
%%%%%%%%%%%%%%%%%%%%%%%%%
for $\tau \simeq \tau_\text{sym} = \omega$, where the relevant symmetry-breaking parameter is given by $\epsilon =(\tau-\omega)/(\tau-\omega^2)$~\cite{Novichkov:2021evw}. In this same basis, the structure of the matrix $M_R$ is unchanged (apart from some signs), thus retaining its Pauli-like block (cf.~\cref{sec:symprot}). Hence, up to unphysical permutations of rows and columns, the full neutrino mass matrix $\mathcal{M}$ has the structure of~\cref{eq:6x6M} in the symmetric limit ($\epsilon \to 0$), as intended.
The Yukawa matrix $Y_e$ is also expected to approach the form given in~\cref{eq:limitYDA1} in the same flavour basis. This structure allows for the generation of charged-lepton mass hierarchies, controlled by the magnitude of $\epsilon$, with a spectrum $\sim 1:\epsilon:\epsilon^2$.
Explicit analytical expressions for the charged-lepton masses in terms of the model parameters are given in~\cref{app:analytics}, for all benchmark models.

In the presence of the gCP symmetry, all parameters are real and some can be taken to be non-negative. Without loss of generality, we choose to keep only the signs of $g_{3,2}$, $\alpha_2$, $\alpha_{3,1}$ and  $\alpha_{3,2}$ as physical.
Thus, the relevant flavour-fit parameters are:
\begin{equation} \label{eq:runparamsA}
    \begin{aligned}
        &\Lambda > 0\,,\quad 
\re \tau \in [-1/2,1/2] \,,\quad   \im \tau > \sqrt{3}/2\,,\\
&g_1 \geq 0\,,\quad  g_2 \geq 0\,,\quad  g_{3,1}\geq 0\,,\quad   g_{3,2} \in \mathbb{R}\,,\\
&\alpha_1\geq 0\,,\quad   \alpha_2\in \mathbb{R}\,,\quad   \alpha_{3,1}\in \mathbb{R}\,,\quad   \alpha_{3,2}\in \mathbb{R}\,.
    \end{aligned}
\end{equation}
There are, in total, 11 real parameters including $\tau$ determining 12 low-energy lepton observables, as well as the 3 HNL masses $M_{1,2,3}$ and the observables described in~\cref{sec:pheno} (see~\cref{fig:ternary_plots,fig:Theta2_plots}).  Of these, 8 are dimensionless superpotential constants
(not imposing gCP would result in the appearance of 4 extra phases). In all considered benchmarks, the dimensionless parameters are scanned in limited ranges to avoid introducing additional hierarchies in the parameters of the model.

\subsubsection*{Model B}

Unlike the previous benchmark, the $N^c$ superfields in this model carry non-zero weights, so that the RH neutrino Majorana mass matrix now reads:
%%%%%%%%%%%%%%%%%%%%
\begin{equation} 
\label{eq:majA4}
    M_R=\Lambda\begin{pmatrix}
    2\left(Y^{(2)}_{\mathbf{3}}\right)_1& -\left(Y^{(2)}_{\mathbf{3}}\right)_3 & -\left(Y^{(2)}_{\mathbf{3}}\right)_2  \\[2mm]
    -\left(Y^{(2)}_{\mathbf{3}}\right)_3& 2\left(Y^{(2)}_{\mathbf{3}}\right)_2 & -\left(Y^{(2)}_{\mathbf{3}}\right)_1  \\[2mm]
    -\left(Y^{(2)}_{\mathbf{3}}\right)_2&-\left(Y^{(2)}_{\mathbf{3}}\right)_1 & 2\left(Y^{(2)}_{\mathbf{3}}\right)_3
\end{pmatrix}\,.
\end{equation}
%%%%%%%%%%%%%%%%%%%%%%%
%
In the appropriate ($ST$-diagonal) flavour basis, one finds the hierarchical structure
%%%%%%%%%%%%%%%%%%%%%%%%%
\begin{equation}
M_R'  \sim \,\begin{pmatrix}
    \epsilon^2& 1 & \epsilon \\
    1 & \epsilon & \epsilon^2 \\
    \epsilon  & \epsilon^2& 1
\end{pmatrix}\,,
\end{equation}
%%%%%%%%%%%%%%%%%%
%
which reproduces the desired form of~\cref{eq:6x6M} in the symmetric limit.
Here, the matrices $Y_D$ and $Y_e$ are obtained from those of model A following the column permutations:
\begin{equation}
Y_{D,e}^{(\text{B})}\,=\,
Y_{D,e}^{(\text{A})}
\begin{pmatrix}
    0 & 0 & 1 \\
    1 & 0 & 0 \\
    0 & 1 & 0 
\end{pmatrix}\,.
\end{equation}
Only the change to $Y_D$ is relevant in practice, since the order of the columns of $Y_e$ is not physical. 
Thus, as in model A, the charged-lepton mass hierarchies are 
given by $\sim (1:\epsilon:\epsilon^2)$.
In the $ST$-diagonal flavour basis, one finds also the hierarchical structure
\begin{equation}
Y_D' \sim \begin{pmatrix}
1 & \epsilon & \epsilon^2 \\
1 & \epsilon & \epsilon^2 \\
1 & \epsilon & \epsilon^2
\end{pmatrix}\,,
\end{equation}
which, in the symmetric limit, directly matches the form of the Dirac block in~\cref{eq:6x6M}. Parameter counting considerations are unchanged with respect to the previous model.

\subsubsection*{Model C}
We now turn to model C, based on the $S_4'$ modular group~\cite{Novichkov:2020eep} and featuring 3 RH$\nu$s. The $N^c$ superfields carry non-zero weights, and the RH neutrino Majorana mass matrix reads 
%%%%%%%%%%%%%%%%%%%%%
 \begin{equation} 
 \label{eq:MajC}
    M_R=\Lambda\begin{pmatrix}
    \frac{2}{\sqrt{3}}\left(Y^{(2)}_{\mathbf{{2}}}\right)_1 & 0 & 0 \\[2mm]
   0&  \left(Y^{(2)}_{\mathbf{{2}}}\right)_2&- \frac{1}{\sqrt{3}}\left(Y^{(2)}_{\mathbf{{2}}}\right)_1\\[2mm]
   0&  - \frac{1}{\sqrt{3}}\left(Y^{(2)}_{\mathbf{{2}}}\right)_1&\left(Y^{(2)}_{\mathbf{{2}}}\right)_2
\end{pmatrix}\,.
\end{equation}
%%%%%%%%%%%%%%%%%%%%%%%%
%
The symmetry-protected mechanism is here realized for $\tau \simeq \tau_\text{sym} = i \infty$,  with symmetry-breaking parameter $\epsilon = q^{1/4} = e^{\pi i \tau/2}$ and $|q|\ll 1$. 
The explicit modular forms in~\cref{app:forms} are already given in the $T$-diagonal basis, appropriate for the hierarchy analysis. One directly finds:
%%%%%%%%%%%%%%%%%%%%%
\begin{equation}
    M_R\sim\begin{pmatrix}
    1 & 0 & 0 \\
   0 &  \epsilon^2& 1\\
  0 & 1&\epsilon^2
\end{pmatrix}\,,
\end{equation}
%%%%%%%%%%%%%%%%%%%%%%
%
in the vicinity of the symmetric point,
showing that the required Pauli-like structure within $M_R$ emerges in the symmetric limit.
For the neutrino Yukawa matrix, one has
\begin{equation}
Y_D =  \begin{pmatrix}
g_1\left(Y^{(3)}_{\mathbf{\hat{3}'}}\right)_1 & g_1\left(Y^{(3)}_{\mathbf{\hat{3}'}}\right)_3 & g_1\left(Y^{(3)}_{\mathbf{\hat{3}'}}\right)_2
\\[2mm]
g_2 \left(Y^{(4)}_{\mathbf{{3}}}\right)_1 & g_2\left(Y^{(4)}_{\mathbf{{3}}}\right)_3 & g_2\left(Y^{(4)}_{\mathbf{{3}}}\right)_2 \\[2mm]
g_{3}\left(Y^{(5)}_{\mathbf{\hat{3}'}}\right)_1 & g_{3}\left(Y^{(5)}_{\mathbf{\hat{3}'}}\right)_3 
& g_{3}\left(Y^{(5)}_{\mathbf{\hat{3}'}}\right)_2
\end{pmatrix}\,.
\end{equation}
which depends only on three superpotential parameters.
Once again, the charged-lepton Yukawa matrix has the same structure as $Y_D$ and reads:
\begin{equation} \label{eq:YeC}
    Y_e =  \begin{pmatrix}
\alpha_1\left(Y^{(3)}_{\mathbf{\hat{3}'}}\right)_1 & \alpha_1\left(Y^{(3)}_{\mathbf{\hat{3}'}}\right)_3 & \alpha_1\left(Y^{(3)}_{\mathbf{\hat{3}'}}\right)_2
\\[2mm]
\alpha_2 \left(Y^{(4)}_{\mathbf{{3}}}\right)_1 & \alpha_2\left(Y^{(4)}_{\mathbf{{3}}}\right)_3 & \alpha_2\left(Y^{(4)}_{\mathbf{{3}}}\right)_2 \\[2mm]
\alpha_{3}\left(Y^{(5)}_{\mathbf{\hat{3}'}}\right)_1 & \alpha_{3}\left(Y^{(5)}_{\mathbf{\hat{3}'}}\right)_3 
& \alpha_{3}\left(Y^{(5)}_{\mathbf{\hat{3}'}}\right)_2
\end{pmatrix}\,.
\end{equation}
Close to the symmetric limit, i.e.~for relatively large $\im\tau$ (see~\cref{foot:imtau}), one finds the limiting structures
\begin{equation} \label{eq:YDeC}
Y_{D,e} \sim\begin{pmatrix}
 \epsilon^3 & \epsilon^2 & 1 \\
 \epsilon^2& \epsilon & \epsilon^3 \\
 \epsilon^3 & \epsilon^2 & 1
\end{pmatrix}\,,
\end{equation}
leading to a symmetry-protected form for $m_D$ and 
to a hierarchical charged-lepton mass spectrum $\sim 1:\epsilon:\epsilon^3$.

In the presence of the gCP symmetry, all parameters are real and some can be taken to be non-negative. Without loss of generality, we choose to keep only the signs of $\alpha_2$ and $\alpha_3$ as physical.
Thus, the relevant parameters are:
\begin{equation} \label{eq:runparamsC}
    \begin{aligned}
        &\Lambda > 0\,,\quad 
\re \tau \in [-1/2,1/2] \,,\quad   \im \tau > \sqrt{3}/2\,,\\
&g_1 \geq 0\,,\quad  g_2 \geq 0\,,\quad  g_{3}\geq 0\,,\quad
\alpha_1\geq 0\,,\quad  \alpha_2\in \mathbb{R}\,,\quad   \alpha_{3}\in \mathbb{R}\,.
    \end{aligned}
\end{equation}
There are, in total, 9 real parameters (including $\tau$), out of which 6 are dimensionless superpotential constants. Not imposing gCP would result in the appearance of 2 extra phases.

\subsubsection*{Model D}

This model is obtained from the previous one by changing the 1D irreps of two of the $L_\alpha$.
Since one still has $N^c \sim (\mathbf{3},+1)$, the RH neutrino Majorana mass matrix is once again given by~\cref{eq:MajC}.
The neutrino and charged-lepton Yukawa matrices $Y_D$ and $Y_e$ include instead an additional parameter each. One has
\begin{equation}
    Y_D =  \begin{pmatrix}
g_1\left(Y^{(3)}_{\mathbf{\hat{3}}}\right)_1 & g_1\left(Y^{(3)}_{\mathbf{\hat{3}}}\right)_3 & g_1\left(Y^{(3)}_{\mathbf{\hat{3}}}\right)_2
\\[2mm]
g_2 \left(Y^{(4)}_{\mathbf{{3}}}\right)_1 & g_2\left(Y^{(4)}_{\mathbf{{3}}}\right)_3 & g_2\left(Y^{(4)}_{\mathbf{{3}}}\right)_2 \\[2mm]
g_{3,1}\left(Y^{(5)}_{\mathbf{\hat{3}},1}\right)_1 + 
g_{3,2}\left(Y^{(5)}_{\mathbf{\hat{3}},2}\right)_1 & g_{3,1}\left(Y^{(5)}_{\mathbf{\hat{3}},1}\right)_3 + g_{3,2}\left(Y^{(5)}_{\mathbf{\hat{3}},2}\right)_3 & g_{3,1}\left(Y^{(5)}_{\mathbf{\hat{3}},1}\right)_2 + g_{3,2}\left(Y^{(5)}_{\mathbf{\hat{3}},2}\right)_2 
\end{pmatrix}\,,
\end{equation}
and the same structure emerges for $Y_e$,
\begin{equation}
    Y_e =  \begin{pmatrix}
\alpha_1\left(Y^{(3)}_{\mathbf{\hat{3}}}\right)_1 & \alpha_1\left(Y^{(3)}_{\mathbf{\hat{3}}}\right)_3 & \alpha_1\left(Y^{(3)}_{\mathbf{\hat{3}}}\right)_2
\\[2mm]
\alpha_2 \left(Y^{(4)}_{\mathbf{{3}}}\right)_1 & \alpha_2\left(Y^{(4)}_{\mathbf{{3}}}\right)_3 & \alpha_2\left(Y^{(4)}_{\mathbf{{3}}}\right)_2 \\[2mm]
\alpha_{3,1}\left(Y^{(5)}_{\mathbf{\hat{3}},1}\right)_1 + 
\alpha_{3,2}\left(Y^{(5)}_{\mathbf{\hat{3}},2}\right)_1 & \alpha_{3,1}\left(Y^{(5)}_{\mathbf{\hat{3}},1}\right)_3 + \alpha_{3,2}\left(Y^{(5)}_{\mathbf{\hat{3}},2}\right)_3 & \alpha_{3,1}\left(Y^{(5)}_{\mathbf{\hat{3}},1}\right)_2 + \alpha_{3,2}\left(Y^{(5)}_{\mathbf{\hat{3}},2}\right)_2 
\end{pmatrix}\,.
\label{eq:YeD}
\end{equation}
Close to the symmetric limit, one finds
\begin{equation}
Y_{D,e} \sim\begin{pmatrix}
 \epsilon & 1  & \epsilon^2  \\
 \epsilon^2& \epsilon & \epsilon^3 \\
 \epsilon & 1  & \epsilon^2 
\end{pmatrix}\,,
\end{equation}
leading once again to a symmetry-protected form for $m_D$ and 
to a hierarchical charged-lepton mass spectrum $\sim 1:\epsilon:\epsilon^3$.
In the presence of gCP and without loss of generality, only the signs of $g_{3,2}$, $\alpha_2$, $\alpha_{3,1}$ and  $\alpha_{3,2}$ are physical. The parameter counting matches that of model A, see~\cref{eq:runparamsA}. There is a total of 11 real parameters (including $\tau$), out of which 8 are dimensionless (once again, lifting gCP leads to 4 extra phases).

\subsection{Inference results}

%%%%%%%%%%%%%%%%%%%%%%
\begin{table}[t!]
\centering
\renewcommand{\arraystretch}{1.2}
\begin{tabular}{l|cc} 
\toprule
Observable & \multicolumn{2}{c}{Best-fit value and $1\sigma$ range} \\ 
\midrule
$m_e / m_\mu$ & \multicolumn{2}{c}{$0.0048 \pm 0.0002$} \\
$m_\mu / m_\tau$ & \multicolumn{2}{c}{$0.0565 \pm 0.0045$} \\ 
\midrule
& NO & IO \\
$\Delta m_{21}^2/(10^{-5}\text{ eV}^2)$ & \multicolumn{2}{c}{$7.49 \pm 0.19$} \\
$|\Delta m_{31(32)}^2|/(10^{-3}\text{ eV}^2)$ & $2.513 \pm 0.020$ & $2.484 \pm 0.020$ \\
$r \equiv \Delta m_{21}^2/|\Delta m_{31(32)}^2|$ & $0.0298
\pm0.0008$ & $0.0301\pm0.0008$\\
$\sin^2\theta_{12}$ & $0.308\pm 0.012$ & $0.308\pm 0.012$ \\
$\sin^2\theta_{13}$ & $0.02215 \pm 0.00057$ & $0.02231 \pm 0.00056$ \\
$\sin^2\theta_{23}$ & $0.510 \pm 0.025$& $0.512 \pm 0.024$ \\
\bottomrule
\end{tabular}
\caption{
Best-fit values and 1$\sigma$ ranges for 
neutrino oscillation parameters obtained from the \texttt{NuFit 6.0} global analysis~\cite{Esteban:2024eli}, and for charged-lepton mass ratios,
given at the scale $2\times 10^{16}$ GeV with the $\tan \beta \equiv v_u/v_d$ averaging
described in~\cite{Feruglio:2017spp}, obtained from Ref.~\cite{Ross:2007az}.
Note that for $\sin^2\theta_{23}$, 
in place of the non-Gaussian one-dimensional projections, we have
considered a Gaussian approximation based on the $3\sigma$ ranges given in~\cite{Esteban:2024eli}.
}
\label{tab:leptondata}
\end{table}
%%%%%%%%%%%%%%%%%%%%%%
%

\begin{figure}[p]
    \hspace{-1.4cm}
    \includegraphics[width=1.15\textwidth]{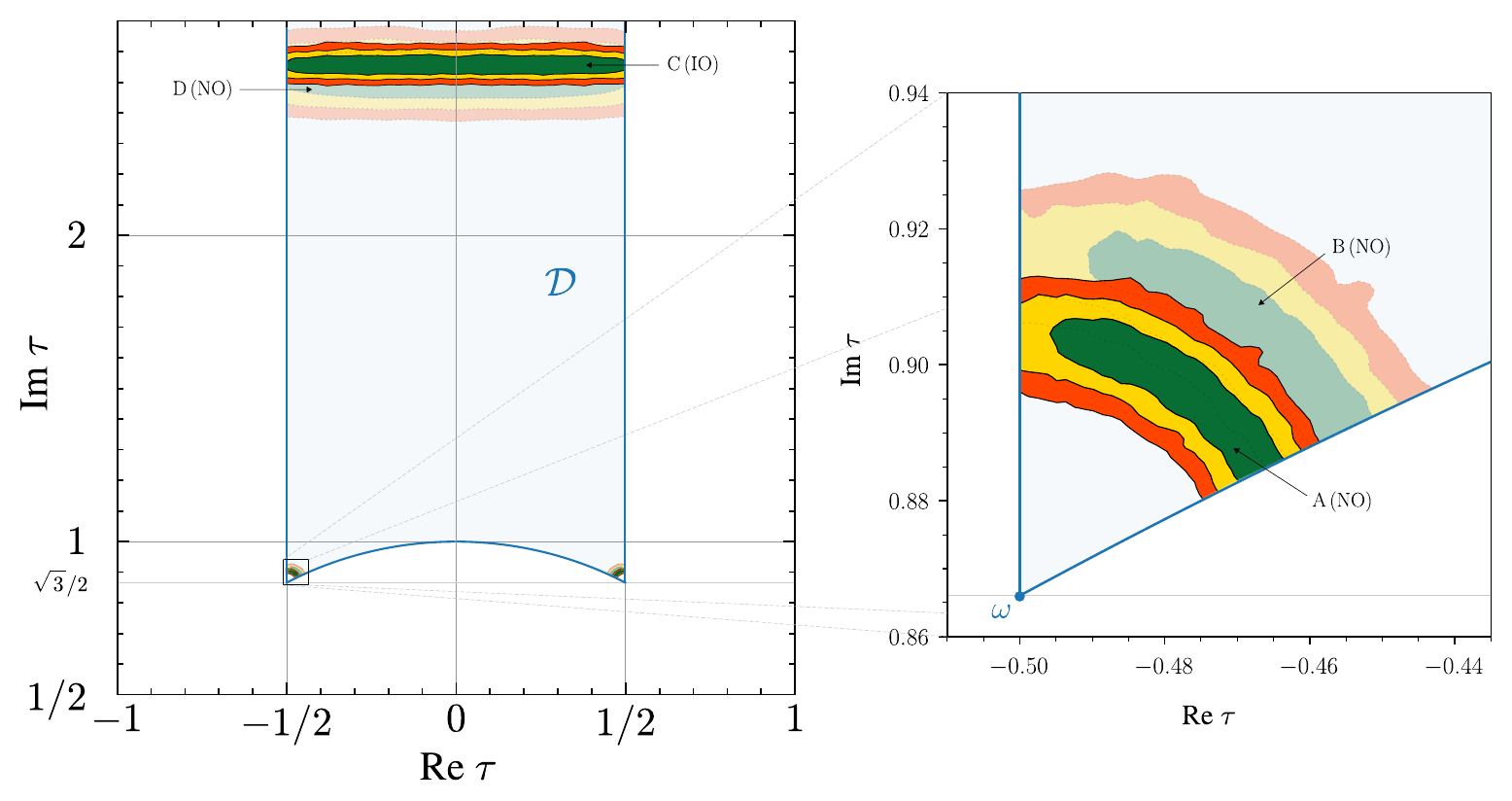}
    \caption{Viable regions for the modulus VEV $\tau$ within the fundamental domain $\mathcal{D}$, for the benchmark modular-symmetry-protected models described in~\cref{sec:modelsubsec}. 
    Green, yellow and red fills correspond to the 1$\sigma$, 2$\sigma$ and 3$\sigma$ credible regions.
    These regions are symmetric under the gCP transformation that flips the sign of $\re \tau$. 
    The panel on the right shows a zoomed-in view near $\tau_\text{sym} = \omega$.}
\label{fig:tau_regions}
\end{figure}

At leading order in the seesaw expansion, the active flavour neutrino Majorana mass matrix $m_\nu$ is diagonalized as
\begin{equation}
\hat{m}_\nu \,\simeq\, U_\nu^T m_\nu U_\nu\,, 
\end{equation}
where $U_\nu$ is a unitary matrix and $\hat{m}_\nu = \diag(m_1, m_2, m_3)$, with $m_{1,2,3}$ denoting the masses of the light active neutrinos. Then, $U = U_e^\dagger U_\nu$ is the unitary Pontecorvo-Maki-Nakagawa-Sakata (PMNS) neutrino mixing matrix, with 
$U_e$ being the unitary matrix arising from the diagonalization of the charged-lepton mass matrix.
In what follows, we adopt the standard numbering of the light neutrinos, having a spectrum with either normal ordering (NO) $m_1 < m_2 < m_3$ or inverted ordering (IO) $m_3<m_1<m_2$, and the standard parameterization
for the PMNS matrix~\cite{Tanabashi:2018oca}:
%%%%%%%%%%%%%%%%%%%%%%%%%%%%
\begin{equation}
\label{eq:PMNS}
U = \begin{pmatrix}
c_{12}c_{13}&s_{12}c_{13}&s_{13}\text{e}^{-i\delta}\\
-s_{12}c_{23}-c_{12}s_{23}s_{13}\text{e}^{i\delta}&c_{12}c_{23}-s_{12}s_{23}s_{13}\text{e}^{i\delta}&s_{23}c_{13}\\
s_{12}s_{23}-c_{12}c_{23}s_{13}\text{e}^{i\delta}&-c_{12}s_{23}-s_{12}c_{23}s_{13}\text{e}^{i\delta}&c_{23}c_{13}
\end{pmatrix}
\begin{pmatrix}
1&0&0\\
0&\text{e}^{i\alpha_{21}/2}&0\\
0&0& \text{e}^{i\alpha_{31}/2}
\end{pmatrix}\,,
\end{equation}
%%%%%%%%%%%%%%%%%%
%
where $c_{ij} \equiv \cos\theta_{ij}$, $s_{ij} \equiv \sin\theta_{ij}$,
$0\leq \delta < 2\pi$ is the Dirac phase,
while $0 \leq \alpha_{21}, \alpha_{31} < 2\pi$ are the two Majorana phases~\cite{Bilenky:1980cx}.
The numerical analysis was conducted by comparing the following six dimensionless observables:
$$\{\sin^2\theta_{12},\,\sin^2\theta_{13},\,\sin^2\theta_{23},\, \Delta m^2_{21}/|\Delta m^2_{31(32)}|, \,m_e/m_\mu, \, m_\mu/m_\tau\}\,,$$
to the respective values in~\cref{tab:leptondata}, where $\Delta m^2_{21} \equiv m_2^2-m_1^2$ and $\Delta m^2_{31(32)} \equiv m_3^{2}-m_{2(1)}^2$. The parameter space was scanned with the use of \texttt{MultiNest}~\cite{Feroz:2013hea}, admitting flat distributions for the priors of the parameter ratios $\beta_i/\beta_j$. The modulus $\tau$ was allowed to vary in the fundamental domain $\mathcal{D}$, in the proximity of $\tau\simeq \omega$ and $\text{Im}\,\tau \gtrsim 2$ for the $A_4$ and $S_4'$ benchmarks, respectively. The four benchmark models are all able to reproduce the lepton data within $1\sigma$, as can be seen from~\cref{tab:results}.
The viable regions for the modulus VEV in each benchmark model are shown in~\Cref{fig:tau_regions}.

Models A, B and D predict a spectrum of light neutrino masses with NO, while model C, which interestingly features the smallest number of free parameters, predicts IO.%
\footnote{A viable region within the parameter space of model D with $\im\tau \sim 3.5$ (predicting NO) is also found, but relies on hierarchical superpotential parameters and is not considered here.}
In models A, B and C, the lightest neutrino is predicted to be almost massless. It is expected to remain so in practice after taking into account the loop corrections described in~\cref{foot:loop}.
Note that, among the predictions, we obtain approximately CP-conserving phases, as a result of the proximity to $\tau_{\text{sym}}=\{\omega,i\infty\}$, which are CP-conserving points.
Deviations from the CP-conserving values $\delta = 0, \pi$  are found to be at the sub-percent level in all four benchmark models.

\begin{figure}[p]
\vskip -2cm
    \centering
    \includegraphics[width=0.44\textwidth]{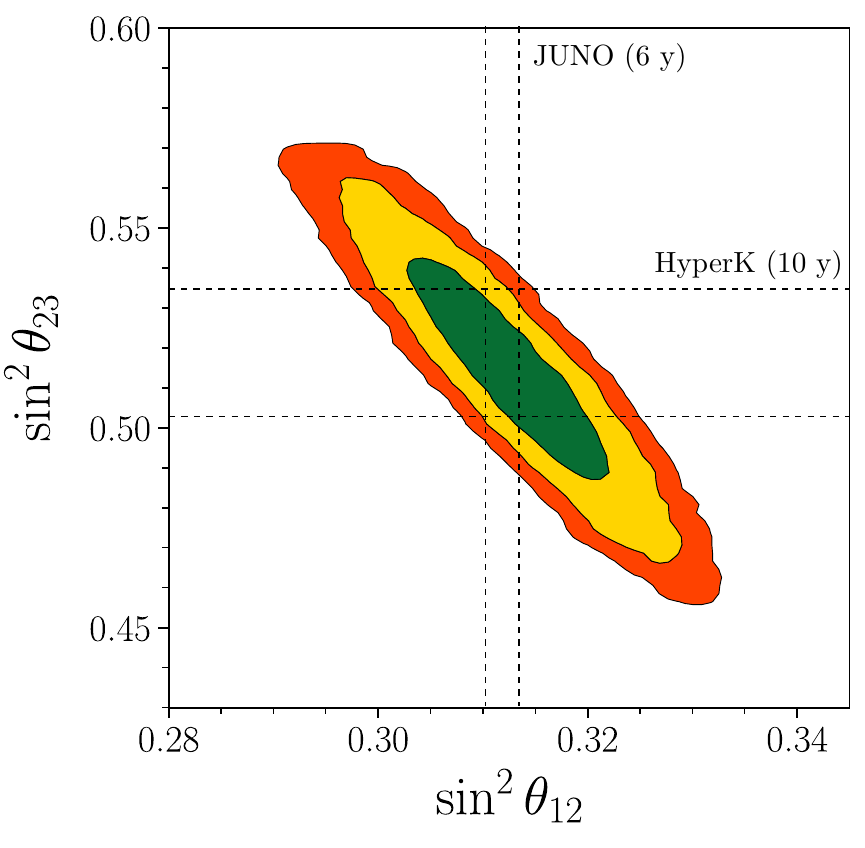}\hspace{5mm}
    \includegraphics[width=0.46\textwidth]{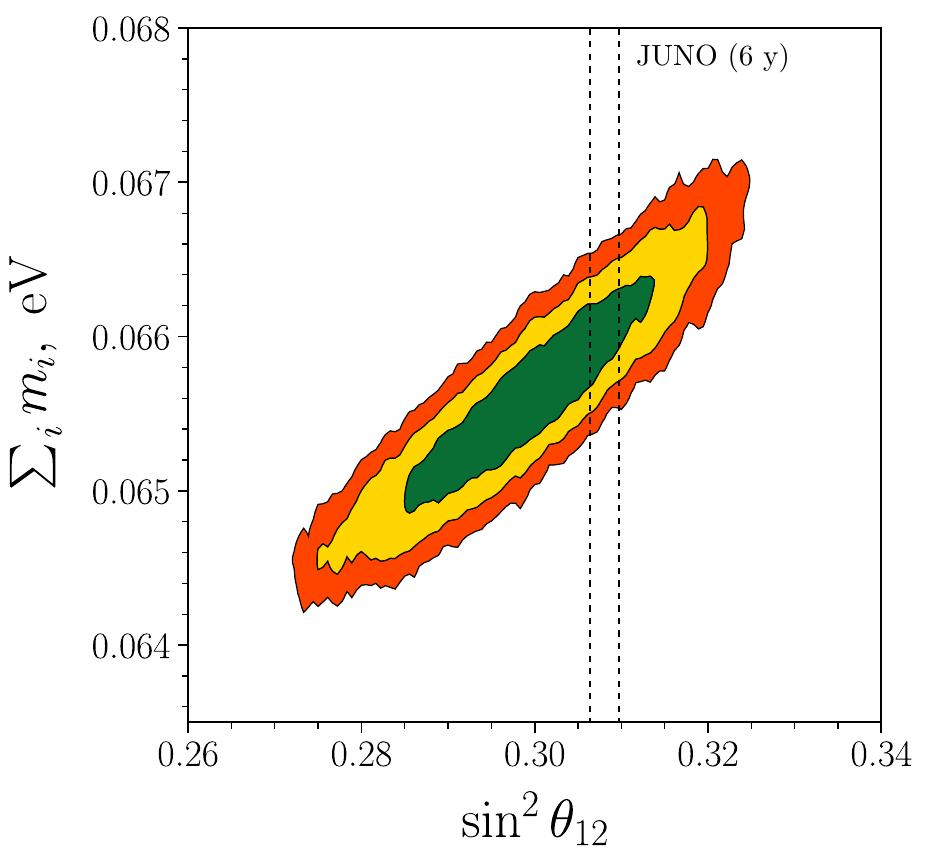}
    \caption{
    Correlation between the solar and atmospheric mixing angles obtained in the assessment of benchmark model C (left) and between the solar angle and the sum of neutrino masses for benchmark model D (right). The colour coding of credible regions is the same as in~\Cref{fig:tau_regions}.
    For comparison, we show the prospective $1\sigma$ sensitivities of future long-baseline and reactor experiments HyperKamiokande~\cite{Hyper-Kamiokande:2025fci} and JUNO~\cite{JUNO:2022mxj}, taking into account the corresponding maximum-likelihood values for $\sin^2\theta_{23}$ and $\sin^2\theta_{12}$, respectively. The predicted values of $\sum_i m_i$ in model D (right) respect the ``aggressive'' cosmological bound identified in Ref.~\cite{Capozzi:2025wyn}.}
\label{fig:corr}
\end{figure}

The magnitudes of the ratios of different free parameters do not span a wide range (with the only exception being model C, where these magnitudes are always $> 100$).
We recall that the parameter ratios $\beta_i/\beta_j$ ($\beta = \alpha, g$) already include the effects of rescaling by the Kähler factors $(2\,\im\,\tau)^{(k_Y^i-k_Y^j)/2}$, where $k_Y$ are the weights of the associated modular forms in the superpotential. In~\cref{tab:results} we report the original, non-rescaled couplings, defined by $ \hat \beta_i =\beta_i(2 \im\tau)^{-k_Y^i/2}$, and their ratios.
It can be argued that the main sources of the required hierarchies in charged-lepton masses come from the chosen modular construction. This can be seen from the division of the obtained mass ratios by the expected power of $\epsilon$, which, as it turns out, is of $\mathcal{O}(1-10)$ in all cases.
We also report, in~\cref{tab:results}, the effective Majorana mass $m_{\beta\beta}$ controlling the rate of neutrinoless double beta decay. Given the smallness of the mass splitting $\Delta M$ and of the HNL mixing $\Theta^2_e$ to the electron flavour in the charged current interaction (see~\cref{sec:pheno}), one estimates the HNL contribution to $m_{\beta\beta}$ to be negligible.
Finally, as shown in~\Cref{fig:corr}, interesting correlations appear between $\sin^2\theta_{12}$ and $\sin^2\theta_{23}$ for model C, and between $\sin^2\theta_{12}$ and $\sum_i m_i$ for model D.

\begin{table}[p]
\small
  \renewcommand{\arraystretch}{1.4}
  \begin{tabular}{lcccc}
  \toprule
Model (ordering)  &  A (NO) & B (NO) & C (IO) & D (NO) \\
\midrule
%%%%%%%%%%%%%%%%%%%%%%%%%%%%%%%%%%%%%
$\re \tau$                            
& $-0.472_{-0.028}^{+0.017}$
& $-0.475_{-0.024}^{+0.036}$
& $\left[-1/2,+1/2\right]$
& $\left[-1/2,+1/2\right]$ \\
%%%%%%%%%%%%%%%%%%%%%%%%%%%%%%%%%%%%%
$\im \tau$                            
& $0.892_{-0.042}^{+0.019}$
& $0.912_{-0.062}^{+0.015}$
& $2.55_{-0.04}^{+0.07}$
& $2.54_{-0.12}^{+0.13}$
\\
%%%%%%%%%%%%%%%%%%%%%%%%%%%%%%%%%%%%%
$\hat\alpha_2 / \hat\alpha_1$
& $-0.286_{-0.106}^{+0.674}$
& $ 0.590_{-0.083}^{+0.125}$ 
& $-0.593_{-0.078}^{+0.102}$ 
& $1.25_{-0.44}^{+0.34}$
\\
%%%%%%%%%%%%%%%%%%%%%%%%%%%%%%%%%%%%%
$\hat\alpha_{3(,1)} / \hat\alpha_1$
& $-0.160_{-0.481}^{+0.145}$ 
& $-0.0502_{-0.1540}^{+0.0501}$ 
& $ 0.0339_{-0.0024}^{+0.0026}$ 
& $0.0531_{-0.0081}^{+0.0062}$
\\
%%%%%%%%%%%%%%%%%%%%%%%%%%%%%%%%%%%%%
$\hat\alpha_{3,2} / \hat\alpha_1$             
& $ 0.0202_{-0.0514}^{+0.0094}$ 
& $-0.0185_{-0.0062}^{+0.0037}$ 
& --- 
& $0.158_{-0.032}^{+0.025}$
\\
%%%%%%%%%%%%%%%%%%%%%%%%%%%%%%%%%%%%%
$\hat g_2 /\hat g_1$
& $0.712_{-0.076}^{+0.052}$ 
& $0.219_{-0.008}^{+0.011}$ 
& $0.0799_{-0.0125}^{+0.0137}$ 
& $7.1_{-2.2}^{+15.0}$
\\
%%%%%%%%%%%%%%%%%%%%%%%%%%%%%%%%%%%%%
$\hat g_{3(,1)} /\hat g_1$                 
& $0.491_{-0.461}^{+0.196}$ 
& $0.336_{-0.161}^{+0.062}$ 
& $13.2_{-1.2}^{+1.1}$
& $0.0985_{-0.0984}^{+0.0197}$
\\
%%%%%%%%%%%%%%%%%%%%%%%%%%%%%%%%%%%%%
$\hat g_{3,2} /\hat g_1$                    
& $0.214_{-0.551}^{+0.119}$ 
& $0.289_{-0.020}^{+0.018}$ 
& --- 
& $-0.21_{-5.60}^{+0.17}$
\\
%%%%%%%%%%%%%%%%%%%%%%%%%%%%%%%%%%%%%
$v_d\,\hat\alpha_1$, GeV                  
& $0.594_{-0.121}^{+0.070}$ 
& $0.363_{-0.055}^{+0.047}$
& $0.176_{-0.007}^{+0.005}$
& $0.161_{-0.013}^{+0.015}$
\\
%%%%%%%%%%%%%%%%%%%%%%%%%%%%%%%%%%%%%
$v_u^2\,\hat g_1^2 / \Lambda$, eV     
& $0.0328_{-0.0070}^{+0.0145}$
& $0.0568_{-0.0120}^{+0.0192}$ 
& $0.00032_{-0.00003}^{+0.00005}$ 
& $0.0104_{-0.0094}^{+0.0053}$
\\
%%%%%%%%%%%%%%%%%%%%%%%%%%%%%%%%%%%%%
\midrule
$m_e / m_\mu$            
& $0.0048_{-0.0005}^{+0.0006}$ 
& $0.0047_{-0.0005}^{+0.0006}$
& $0.0048_{-0.0006}^{+0.0005}$
& $0.0049_{-0.0006}^{+0.0004}$
\\
%%%%%%%%%%%%%%%%%%%%%%%%%%%%%%%%%%%%%
$m_\mu / m_\tau$     
& $0.0560_{-0.0114}^{+0.0115}$ 
& $0.0571_{-0.0121}^{+0.0114}$ 
& $0.0577_{-0.0134}^{+0.0109}$ 
& $0.0580_{-0.0113}^{+0.0108}$
\\
%%%%%%%%%%%%%%%%%%%%%%%%%%%%%%%%%%%%%
$r$                         
& $0.0297_{-0.0021}^{+0.0022}$ 
& $0.0297_{-0.0021}^{+0.0023}$
& $0.0300_{-0.0021}^{+0.0023}$
& $0.0299_{-0.0021}^{+0.0019}$
\\
%%%%%%%%%%%%%%%%%%%%%%%%%%%%%%%%%%%%%
$\sin^2 \theta_{12}$             
& $0.307_{-0.030}^{+0.034}$ 
& $0.308_{-0.030}^{+0.032}$
& $0.312_{-0.018}^{+0.019}$
& $0.308_{-0.033}^{+0.024}$
\\
%%%%%%%%%%%%%%%%%%%%%%%%%%%%%%%%%%%%%
$\sin^2 \theta_{13}$          
& $0.0220_{-0.0014}^{+0.0017}$ 
& $0.0221_{-0.0014}^{+0.0014}$
& $0.0222_{-0.0014}^{+0.0016}$ 
& $0.0220_{-0.0013}^{+0.0017}$
\\
%%%%%%%%%%%%%%%%%%%%%%%%%%%%%%%%%%%%%
$\sin^2 \theta_{23}$              
& $0.506_{-0.065}^{+0.067}$ 
& $0.507_{-0.062}^{+0.065}$ 
& $0.519_{-0.058}^{+0.049}$
& $0.507_{-0.051}^{+0.070}$ 
\\
%%%%%%%%%%%%%%%%%%%%%%%%%%%%%%%%%%%%%
\midrule
$m_1$, eV
& $< 10^{-4}$
& $< 10^{-4}$
& $0.0491_{-0.0002}^{+0.0002}$ 
& $0.0054_{-0.0010}^{+0.0009}$
\\
%%%%%%%%%%%%%%%%%%%%%%%%%%%%%%%%%%%%%
$m_2$, eV 
& $0.00864_{-0.00029}^{+0.00028}$ 
& $0.00865_{-0.00028}^{+0.00029}$ 
& $0.0499_{-0.0002}^{+0.0001}$ 
& $0.0102_{-0.0006}^{+0.0005}$
\\
%%%%%%%%%%%%%%%%%%%%%%%%%%%%%%%%%%%%%
$m_3$, eV
& $0.0501_{-0.0002}^{+0.0001}$ 
& $0.0501_{-0.0002}^{+0.0002}$
& $< 10^{-4}$
& $0.0504_{-0.0002}^{+0.0002}$
\\
%%%%%%%%%%%%%%%%%%%%%%%%%%%%%%%%%%%%%
$\Sigma_i m_i$, eV 
& $0.0588_{-0.0001}^{+0.0001}$ 
& $0.0588_{-0.0001}^{+0.0001}$ 
& $0.0990_{-0.0004}^{+0.0003}$ 
& $0.0660_{-0.0015}^{+0.0015}$ 
\\
%%%%%%%%%%%%%%%%%%%%%%%%%%%%%%%%%%%%%
$m_{\beta\beta}$, meV
&  $1.49_{-0.28}^{+0.31}$
&  $1.49_{-0.28}^{+0.27}$
&  $17.8_{-1.8}^{+1.7}$
&  $1.69_{-0.11}^{+0.13}$ 
\\
%%%%%%%%%%%%%%%%%%%%%%%%%%%%%%%%%%%%%
$\delta$
& $\simeq 0,\,\pi$
& $\simeq 0$
& $\simeq 0$
& $\simeq \pi$
\\
%%%%%%%%%%%%%%%%%%%%%%%%%%%%%%%%%%%%%
$\alpha_{21(23)}$
& $\simeq \pi$ 
& $\simeq \pi$ 
& $\simeq 0$
& $\simeq \pi$
\\
%%%%%%%%%%%%%%%%%%%%%%%%%%%%%%%%%%%%%
$\alpha_{31}$
& --- 
& --- 
& ---
&  $\simeq 0$
\\
%%%%%%%%%%%%%%%%%%%%%%%%%%%%%%%%%%%%%
$\Delta M$
& $0.0416_{-0.0018}^{+0.0019}$ eV
& $\left(0.149_{-0.034}^{+0.051}\right) M$
& $\left(0.003_{-0.003}^{+0.015}\right) M$
& $\left(0.010_{-0.010}^{+0.014}\right) M$
\\
%%%%%%%%%%%%%%%%%%%%%%%%%%%%%%%%%%%%%
\midrule
$|\epsilon(\tau)|$
& $0.0218_{-0.0046}^{+0.0047}$ 
& $0.0292_{-0.0067}^{+0.0076}$ 
& $0.0182_{-0.0018}^{+0.0013}$
& $0.0184_{-0.0034}^{+0.0039}$
\\
%%%%%%%%%%%%%%%%%%%%%%%%%%%%%%%%%%%%%
$(m_e / m_\tau) / |\epsilon|^{2,3}$
& $0.563_{-0.111}^{+0.167}$ 
& $0.317_{-0.088}^{+0.123}$ 
& $45.8_{-3.1}^{+3.1}$
& $44.9_{-16.0}^{+19.4}$ 
\\
%%%%%%%%%%%%%%%%%%%%%%%%%%%%%%%%%%%%%
$(m_\mu / m_\tau) / |\epsilon|$
& $2.57_{-0.08}^{+0.05}$ 
& $1.96_{-0.24}^{+0.20}$
& $3.16_{-0.52}^{+0.39}$
& $3.14_{-0.37}^{+0.20}$
\\
%%%%%%%%%%%%%%%%%%%%%%%%%%%%%%%%%%%%%
\midrule
$\min N \sigma$            
& 0.412
& 0.411
& 0.548
& 0.552
\\
%%%%%%%%%%%%%%%%%%%%%%%%%%%%%%%%%%%%%
max par.~ratio         
& 49.6
& 54.1
& 165
& 71.8
\\
%%%%%%%%%%%%%%%%%%%%%%%%%%%%%%%%%%%%%
min(max par.~ratio)
& 32.1
& 40.5
& 133
& 55.2
\\
\bottomrule
  \end{tabular}
  \caption{Central values and limits of the 3$\sigma$ credible regions for the parameters and observables in each of the benchmark models of~\cref{tab:benchmarks}. Also shown are the comparison of charged-lepton mass ratios to the appropriate power of $|\epsilon|$, as well as the largest ratio of superpotential parameters (max par.~ratio) and $N\sigma$, the root of a Gaussian $\chi^2$, at the point of maximum likelihood. In the last line, we also show the lowest value that the largest parameter ratio can attain within the 3$\sigma$ credible region.
  Here, $\epsilon = (\tau-\omega)/(\tau-\omega^2)$ for cases A and B and $\epsilon = e^{\pi i \tau/2}$ for cases C and D. 
  In cases with a nearly massless neutrino, we report the single relevant Majorana phase: $\alpha_{23} \equiv \alpha_{21} - \alpha_{31}$ for NO and $\alpha_{21}$ for IO.
  }
  \label{tab:results}
\end{table}

%\vfill
%\clearpage
%%%%%%%%%%%%%%%%%%%%%%%%%%%%%%%%%%%%%%%%%%%%%%
\section{Heavy neutral lepton phenomenology}
\label{sec:pheno}
%%%%%%%%%%%%%%%%%%%%%%%%%%%%%%%%%%%%%%%%%%%%%%

At first order in the seesaw expansion, the heavy Majorana neutrinos that result from the full diagonalization of the mass terms in~\cref{eq:Lmnu} mix with the active ones. In the basis 
in which the charged-lepton Yukawa and the RH neutrino mass matrices are diagonal, the mixing relation reads:
\begin{equation}
    \nu_{l L} \,\simeq\, U_{l a}\, \nu_{a L} +   \frac{v_u \hat{Y}_{l j}}{M_j}\, N_{j L}\,,
\end{equation}
where here we intend $\nu_{l L}$ as the LH active neutrino fields of flavour $l = e,\,\mu,\,\tau$ that enter the SM electroweak interaction Lagrangian in the chosen basis (not to be confused with the previously-used states in the non-diagonal basis); $\hat{Y} = U^\dagger_e Y^*_D V^*$ are the Dirac-type neutrino Yukawas in the this basis, $Y^*_D$ being the neutrino Yukawa coupling in~\cref{eq:Lseesaw},
$V$ a unitary matrix diagonalizing the RH neutrino Majorana mass term and
$U^\dagger_e$ one of the two unitary matrices diagonalizing the charged-lepton mass matrix $U= U^\dagger_e U_\nu$; $\nu_{a L}$ and $N_{j L}$ are respectively the LH projections of the Majorana fields describing light and heavy neutrinos with masses $m_a$, $a=1, 2,3$ and $M_j$, $j = 1,2,3$. At leading order in the seesaw expansion, $M_{1,2,3}$  are the singular values of the matrix $M_N\simeq M_R$, see~\cref{eq:Mmass}. Thanks to the mixing relation above, the heavy Majorana neutrinos $N_j$ inherit a coupling to the electroweak sector via Charged Current (CC) and Neutral Current (NC) interactions:
%%%%%%%%%%%%%%%%%%%%%%%%%%%%%%%%%%%%%%%%%%%
\begin{align}
 \mathcal{L}_\text{CC}^N&\,\simeq\, -\frac{g}{\sqrt{2}}\, \Theta_{l j}\,
\overline{l_L}\,\,\slashed{W}\,N_{j L}
\,+\,{\rm h.c.}\,,
\label{eq:NCC}\\
\mathcal{L}_\text{NC}^N &\,\simeq\, -\frac{g}{2 c_{w}}\,\Theta_{l j}\,
\overline{\nu_{l L}}\,\,\slashed{Z}\,N_{j L}
\,+\,{\rm h.c.}\,,
\label{eq:NNC}
\end{align}
%%%%%%%%%%%%%%%%%%%%%%%%%%%%%%%%%%%%%%
%
with $\Theta_{l j} \equiv v_u \hat{Y}_{l j}/M_j$.
Here, $g$ is the $\text{SU}(2)_L$ gauge coupling; 
$c_w \equiv \cos\theta_w$, $\theta_w$ being the weak mixing angle; $l_L$ is the LH charged-lepton field of flavour $l = e,\,\mu,\,\tau$.%
\footnote{Through the same mixing relation, the heavy Majorana neutrinos also acquire a coupling to the flavour neutrinos and the Higgs bosons.}

\vskip 2mm
The fact that the heavy Majorana neutrinos $N_j$ couple to the weak bosons implies extensive testable phenomenology, shared with the broader class of HNLs~\cite{Leung:1983ix, Gronau:1984ct, Han:2006ip, delAguila:2007qnc, Gorbunov:2007ak, delAguila:2008cj, delAguila:2008hw, Atre:2009rg, Ibarra:2010xw,Ibarra:2011xn, Dinh:2012bp, Cely:2012bz, Penedo:2017knr, Bondarenko:2018ptm, Bolton:2019pcu, Urquia-Calderon:2022ufc}. If the heavy Majorana neutrinos have masses at or below the electroweak scale,
and their mixing is sufficiently large, they can be copiously produced at colliders or beam-line facilities, and eventually be identified via the subsequent decay (either prompt or displaced) into charged particles (see, e.g.,~\cite{Abdullahi:2022jlv, Antel:2023hkf} and references therein). In these kinds of HNL searches, the phenomenologically-relevant parameters are combinations of $|\Theta_{l j}|^2$. To demonstrate the potential of future HNL searches in testing the benchmark models considered in this work, we focus on the following two key combinations:
\begin{equation}
\Theta^2_l \equiv \sum_{j=1}^3|\Theta_{l j}|^2\qquad\text{and}\qquad
\Theta^2 \equiv \sum_{l = e,\,\mu,\,\tau}\Theta_l^2 \,,
\end{equation}
with the first parameter entering the overall contribution of a given individual flavour to HNL signals, while the second characterizes the total mixing irrespective of the flavour and the specific heavy neutrino state.

As discussed in~\cref{sec:framework}, 
in the modular-symmetry-protected low-scale seesaw scenario considered by us,
the magnitude of the neutrino Yukawa couplings is enhanced with respect to the 
case of the standard seesaw one by a factor which is inversely 
proportional to a power of the small parameter $\epsilon$ 
which determines the charged-lepton mass hierarchies.
The degree of enhancement varies with 
the model. Correspondingly, the values of the heavy Majorana neutrino 
couplings $\Theta^2_{l}$ and $\Theta^2$ are 
also enhanced. This opens up the possibility of testing the  
low-scale seesaw scenario by observing 
the associated heavy Majorana neutrinos and by measuring, e.g., 
 $\Theta^2_{l}$.

\begin{figure}[p]
    \centering
    \includegraphics[width=0.49\linewidth]{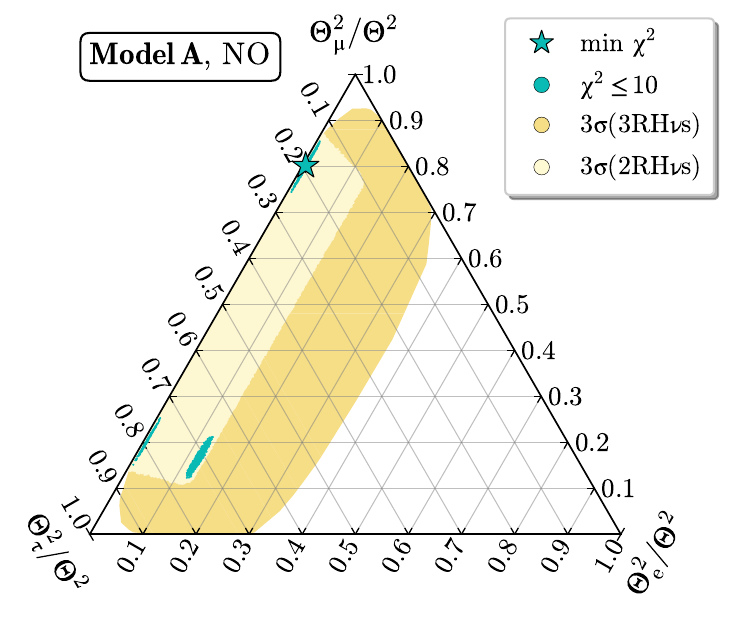}
\includegraphics[width=0.49\linewidth]{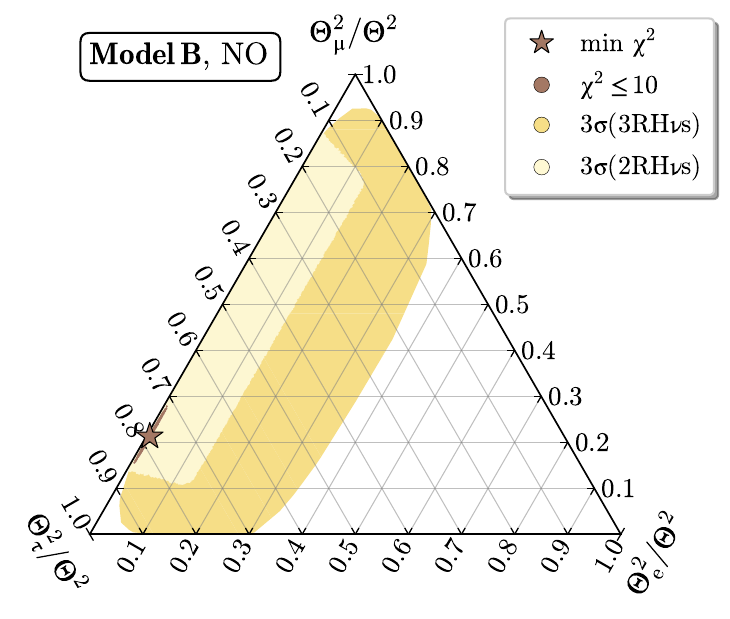}\\
    \includegraphics[width=0.49\linewidth]{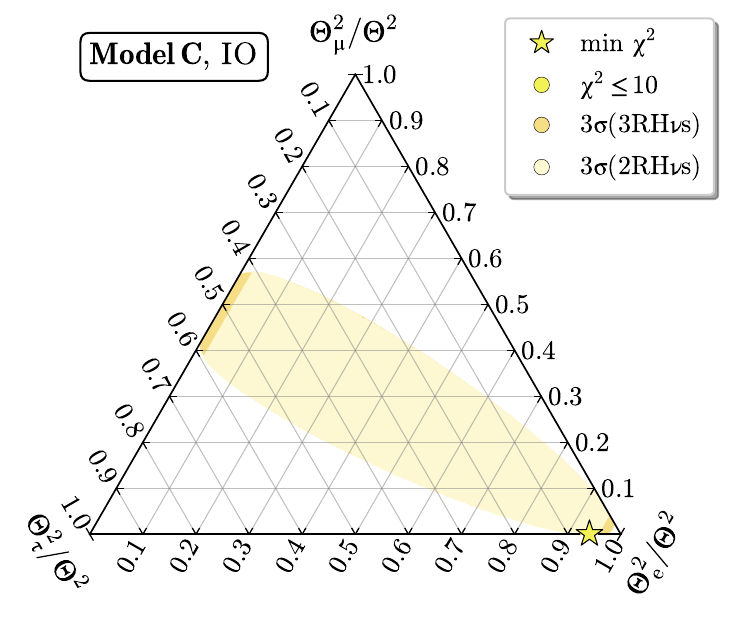}
\includegraphics[width=0.49\linewidth]{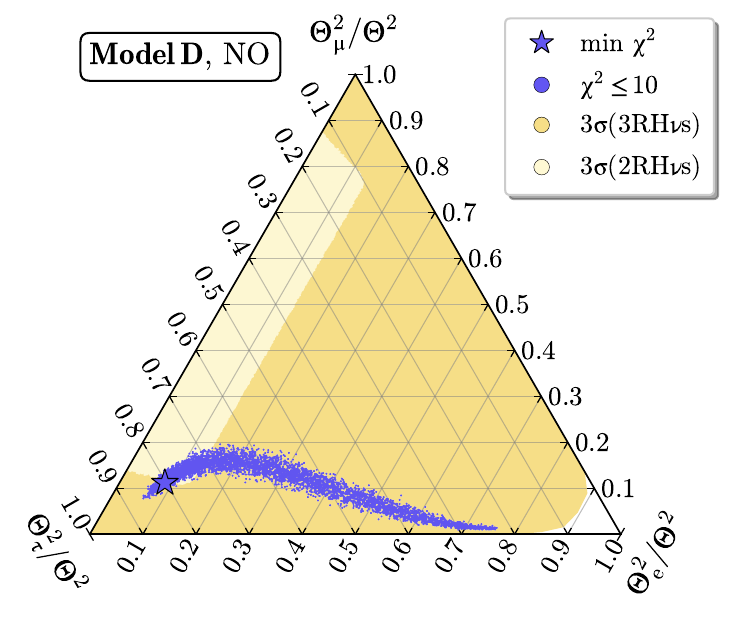}
    \caption{The ratios $\Theta^2_e/\Theta^2$\,--\,$\Theta^2_\mu/\Theta^2$\,--\,$\Theta^2_\tau/\Theta^2$ associated to the considered models. The coloured points -- top-left panel in turquoise for model A, top-right in brown for model B, bottom-left in yellow for model C, and bottom-right in blue for model D -- are those for which $\chi^2 \leq 10$, with the stars marking the point of maximum posterior probability (minimum Gaussian $\chi^2$). The orange regions correspond to the full parameter space of the type-I seesaw, in the cases with either two (lighter colour) or three (darker colour) RH$\nu$s, with the oscillation data varied within the $3\sigma$ regions obtained in the \texttt{NuFit 6.0} global analysis~\cite{Esteban:2024eli}. To obtain the region associated to generic type-I scenario with 3 RH$\nu$s, the lightest neutrino mass is varied randomly in the range allowed by the corresponding model when compared against oscillation data.
    }
\label{fig:ternary_plots}
\end{figure}

We show in the ternary plots of~\cref{fig:ternary_plots} the values of the ratios $\Theta^2_e/\Theta^2$\,--\,$\Theta^2_\mu/\Theta^2$\,--\,$\Theta^2_\tau/\Theta^2$ for the points of our inference analysis after imposing the cut $\chi^2 \leq 10$, and mark with a star the point corresponding to the minimum of this Gaussian $\chi^2$. Together with the points, we show also the region associated to the general type-I seesaw scenario with either two or three RH$\nu$s, by allowing the oscillation data to vary within the 3$\sigma$ ranges given by the \texttt{NuFit 6.0} global analysis~\cite{Esteban:2024eli} (see also~\cref{tab:leptondata}), and the mass of the lightest neutrino fixed according to the results of our inference analysis. For the benchmark models A, B and C, we obtain nearly point-like regions in distinct areas of the ternary parameter space. For model D we obtain a relatively larger area, which nevertheless corresponds to a quite constrained region compared to the full one associated to the type-I seesaw with three RH$\nu$s. These results showcase the potential to distinguish each of the (very predictive) benchmark models based on their HNL flavour content.

We report below the values of the HNL  squared coupling $\Theta^2_{l}$ to each specific flavour $l=e,\,\mu,\,\tau$ that we obtain in the considered models at the points of minimum $\chi^2$ and for $\Mav \simeq 1\,\text{GeV}$, where $\Mav \equiv \sum_{j=1,2,3} M_j/3$. To highlight the hierarchy between $\Theta_e^2$, $\Theta_\mu^2$ and $\Theta_\tau^2$, we present the numbers on each line in descending order.

\begin{center}
\begin{minipage}{0.7\textwidth} 
\begin{itemize}
    \item[\textbf{A:}] $\Theta^2_\mu = 4.06\times 10^{-10}$, $\Theta^2_\tau = 9.76\times 10^{-11}$, $\Theta_e^2 =  2.90\times 10^{-12}$
    \item[\textbf{B:}] $\Theta^2_\tau = 2.55\times 10^{-10}$, $\Theta^2_\mu = 6.96\times 10^{-11}$, $\Theta_e^2 =  2.12\times 10^{-12}$
    \item[\textbf{C:}]  $\Theta_e^2 =  5.13\times 10^{-7~}$, $\Theta^2_\tau = 3.20\times 10^{-8~}$, $\Theta^2_\mu = 4.94\times 10^{-10}$ 
    \item[\textbf{D:}]  $\Theta^2_\tau = 3.83\times 10^{-9~}$, $\Theta^2_\mu = 5.34\times 10^{-10}$, $\Theta_e^2 =  4.08\times 10^{-10}$
\end{itemize}
\end{minipage}
\end{center}
The numbers above make clear that the considered models all lead to distinct predictions in terms of hierarchy and magnitudes of the squared couplings $\Theta_e^2$, $\Theta_\mu^2$ and $\Theta_\tau^2$.\footnote{We note that in the case of Model A, there exist two other separate regions of the ternary parameter space satisfying $\chi^2\leq 10$ that correspond to a flavour hierarchy of the type $\Theta_\tau^2 \gg \Theta_\mu^2> \Theta^2_e$, see the upper-left panel of~\cref{fig:ternary_plots}. Also, for Model D, we find points that satisfy $\Theta_e^2 > \Theta_\tau^2 \gg \Theta_\mu^2$, see the bottom-right panel of~\cref{fig:ternary_plots}.}

\begin{figure}[p]
    \centering
    \includegraphics[width=0.49\linewidth]{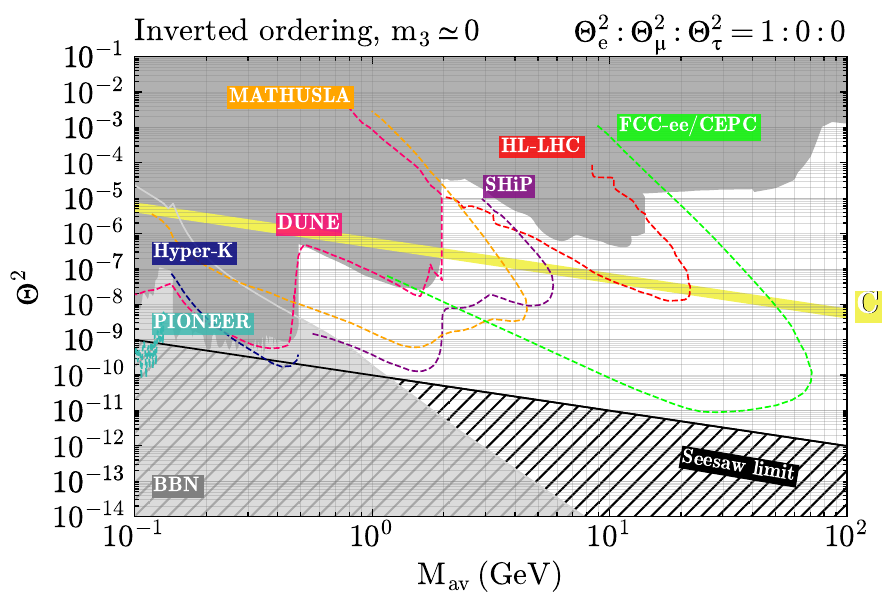}
        \includegraphics[width=0.49\linewidth]{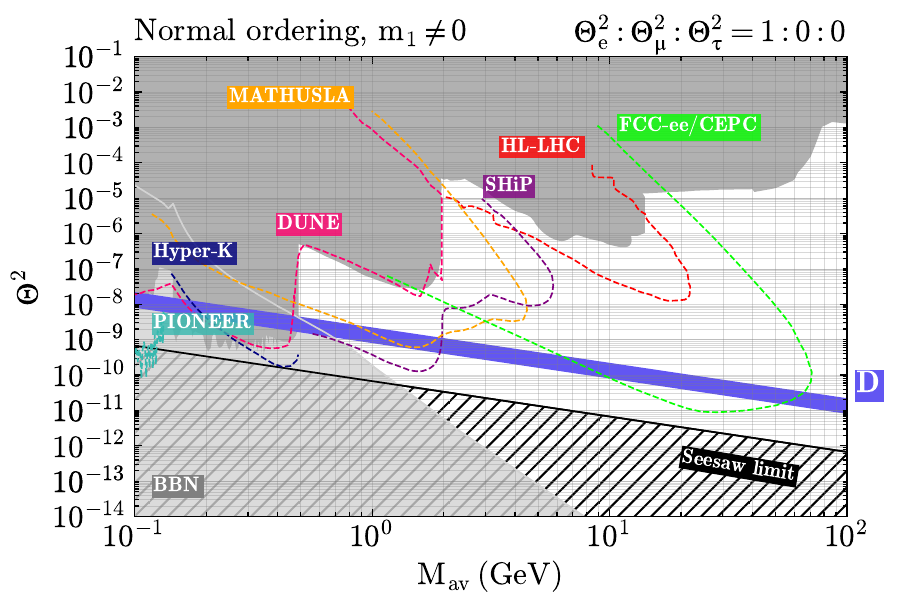}\\
    \includegraphics[width=0.49\linewidth]{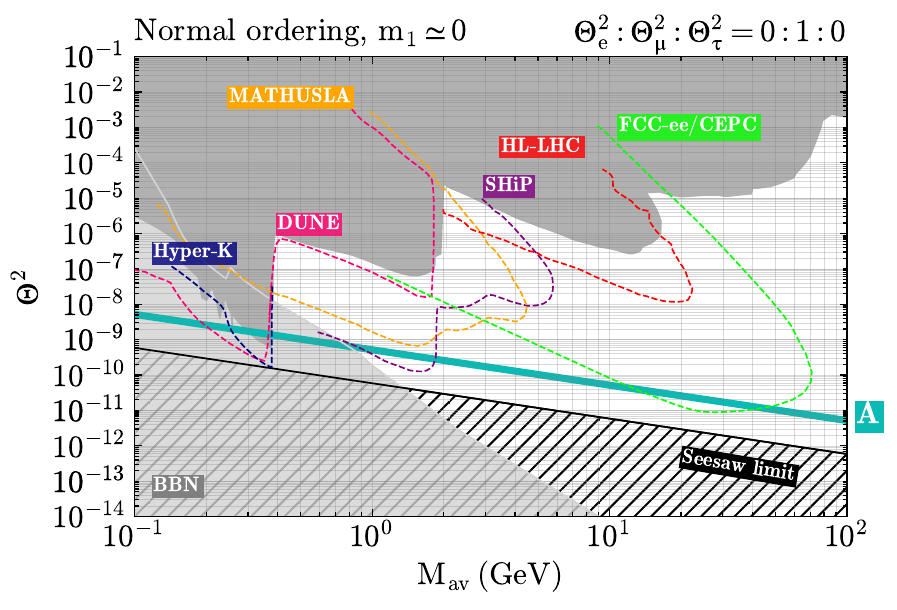}\\
\includegraphics[width=0.49\linewidth]{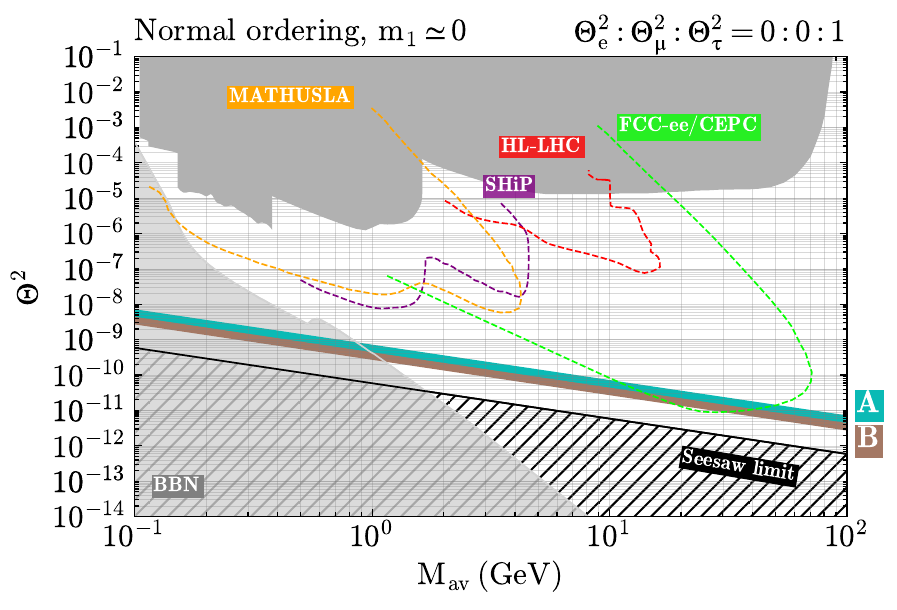}
\includegraphics[width=0.49\linewidth]{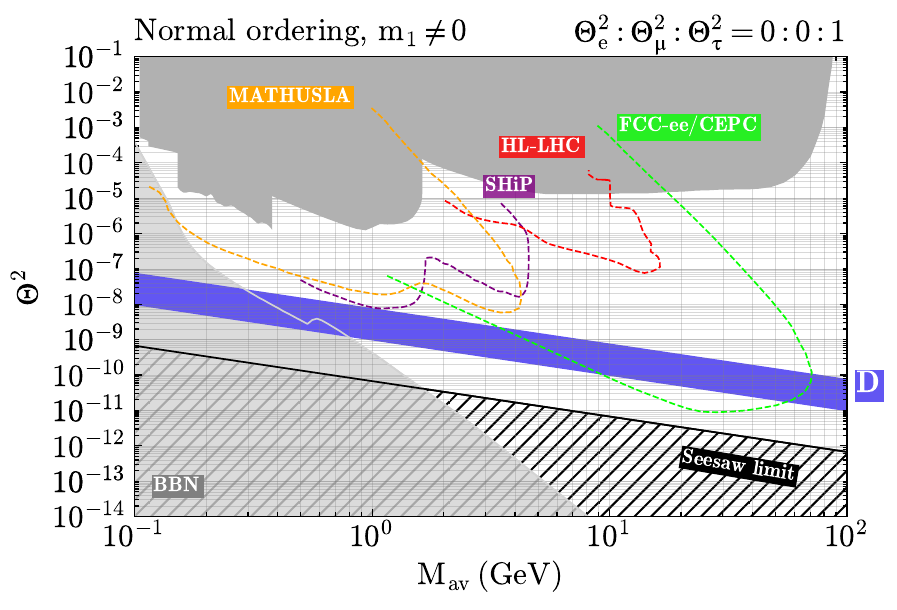}
    \caption{The parameter space in the $\Theta^2$\,--\,$\Mav$ plane of the type-I seesaw scenario for the discussed benchmark models: models C and D with $\Theta_e^2$-dominance in the top panels, model A with $\Theta_\mu^2$-dominance in the middle panel, models A, B and D with $\Theta^2_\tau$-dominance in the bottom panels. The black solid curve represents the seesaw limit. The darker gray regions are excluded by several experiments on HNL production via meson decays (PS191~\cite{Bernardi:1985ny, Bernardi:1987ek}, BEBC~\cite{Barouki:2022bkt}, PIENU~\cite{PIENU:2017wbj}, E949~\cite{E949:2014gsn}, NA62~\cite{NA62:2020mcv, NA62:2021bji}, T2K~\cite{T2K:2019jwa}, NuTeV~\cite{NuTeV:1999kej}, MicroBooNE~\cite{MicroBooNE:2022ctm, MicroBooNE:2023eef} CHARM~\cite{CHARM:1985nku, Boiarska:2021yho}, searches at KEK~\cite{Hayano:1982wu}), tau lepton decays (BELLE~\cite{Belle:2013ytx, Belle:2022tfo, Belle:2024wyk}), and at colliders (DELPHI~\cite{DELPHI:1996qcc}, CMS~\cite{CMS:2022fut, CMS:2023jqi, CMS:2024xdq}, ATLAS~\cite{ATLAS:2019kpx, ATLAS:2022atq, Tastet:2021vwp}). The lighter gray region is excluded by BBN~\cite{Sabti:2020yrt, Boyarsky:2020dzc}. The dashed curves represent the sensitivities of the upcoming, planned and proposed experiments PIONEER~\cite{PIONEER:2022yag} (cyan), Hyper-K~\cite{T2K:2019jwa} (blue), DUNE~\cite{Breitbach:2021gvv} (pink), MATHUSLA~\cite{MATHUSLA:2020uve} (orange), SHiP~\cite{SHiP:2018xqw} (purple), and searches at HL-LHC~\cite{Drewes:2019fou} (red), FCC-ee~\cite{Blondel:2022qqo} and CEPC~\cite{CEPCStudyGroup:2018ghi} (green). Current constraints and future sensitivities are only indicative, as they are given for $\Theta_e^2 : \Theta^2_\mu:\Theta^2_\tau = 1:0:0$ (upper panel), $0:1:0$ (middle panel) and $0:0:1$ (lower panels) -- such ratios hold only approximately in the scenarios considered here -- and in the case of a single HNL.
    }
\label{fig:Theta2_plots}
\end{figure}

We show in~\cref{fig:Theta2_plots} the parameter space in the $\Theta^2$\,--\,$\Mav$ plane of the type-I seesaw scenario for the discussed models with modular symmetry. We have fixed $\tan\beta = v_u / v_d = 5$ in all four benchmarks, with $v \equiv (v_u^2 + v_d^2)^{1/2} \simeq 174\,\text{GeV}$ and $v_u \simeq 0.98\,v$. The solid black curve represents the seesaw limit computed as $\Theta^2 = \sum_{a=1,2,3} m_a/\Mav$. The dark gray region is excluded by past HNL searches~\cite{Bernardi:1985ny, Bernardi:1987ek, Barouki:2022bkt, PIENU:2017wbj, E949:2014gsn, NA62:2020mcv, NA62:2021bji, T2K:2019jwa, NuTeV:1999kej, MicroBooNE:2022ctm, MicroBooNE:2023eef, CHARM:1985nku, Boiarska:2021yho, Hayano:1982wu, Belle:2013ytx, Belle:2022tfo, Belle:2024wyk, DELPHI:1996qcc, CMS:2022fut, CMS:2023jqi, CMS:2024xdq, ATLAS:2019kpx, ATLAS:2022atq, Tastet:2021vwp}, while the lighter gray one by Big Bang nucleosynthesis (BBN)~\cite{Sabti:2020yrt, Boyarsky:2020dzc}. The dashed curves represent the forecasted sensitivities of various planned and proposed experiments \cite{PIONEER:2022yag, T2K:2019jwa, Breitbach:2021gvv, MATHUSLA:2020uve, SHiP:2018xqw, Drewes:2019fou, Blondel:2022qqo, CEPCStudyGroup:2018ghi}. Given the predicted hierarchies among the couplings $\Theta_e^2$, $\Theta^2_\mu$ and $\Theta^2_\tau$, we show the present limits and future sensitivities on $\Theta^2$ assuming either $\Theta_e^2 : \Theta_\mu^2 : \Theta_\tau^2 = 1:0:0$ (Models C and D), $\Theta_e^2 : \Theta_\mu^2 : \Theta_\tau^2 = 0:1:0$ (Model A), $\Theta_e^2 : \Theta_\mu^2 : \Theta_\tau^2 = 0:0:1$ (Models A, B and D). To obtain the parameter space associated to Model A in the $\Theta^2$\,--\,$\Mav$ plane, we impose a cut $\Theta_\mu^2/\Theta^2>0.5$ ($<0.5$) to select points that approximately satisfy $\Theta_\mu^2 \gg \Theta_\tau^2, \,\Theta_e^2$ ($\Theta_\tau^2 \gg \Theta_\mu^2, \,\Theta_e^2$). Analogously, for Model D, we impose $\Theta_e^2/\Theta^2>0.5$ ($<0.5$) to select points satisfying approximately $\Theta_e^2 \gg \Theta_\tau^2, \,\Theta_\mu^2$ ($\Theta_\tau^2 \gg \Theta_\mu^2 ,\, \Theta_e^2$). We emphasize that the assumed ratios hold only approximately in the scenario considered by us, and, because of this, the presented limits and sensitivities should be taken as indicative. To be more precise, we should recast these limits and sensitivities assuming the exact flavour hierarchies predicted by our models, but such an analysis is beyond the scope of the present study. 

It stems from~\cref{fig:Theta2_plots} that the proposed mechanism lifts the allowed parameter space from the seesaw limit
into the testable region, allowing to probe the flavour structure illustrated in~\cref{fig:ternary_plots}.
Furthermore, the expected ranges of values for $\Theta^2$ represent distinct signatures of each benchmark, which, depending on $\Mav$, can be probed at different future experiments.
A striking result is the extra enhancement of HNL couplings within model C, which stems from the hierarchies between the $\hat{g}_i$ parameters in the viable model region (see~\cref{tab:results}). Such hierarchies lead, in that case, to a departure from the power structure of~\cref{eq:YDeC} that provides an additional protection of the L symmetry.

\vskip 2mm
The CC and NC couplings in~\cref{eq:NCC,eq:NNC} generate charged-lepton flavour violating (cLFV) processes at one-loop, with the exchange of virtual
HNLs~\cite{Petcov:1976ff,Bilenky:1977du}. These processes provide an alternative indirect method to probe HNLs. Upcoming experiments looking for cLFV processes involving muons, namely MEG II~\cite{MEGII:2018kmf} on $\mu\to e\gamma$ decay, Mu3e Project~\cite{Arndt:2009} on $\mu \to eee$ decay, Mu2e~\cite{Bartoszek:2015} and COMET~\cite{Abramishvili:2020} (PRISM/PRIME~\cite{Barlow:2011zza}) on $\mu$\,--\,$e$ conversion in aluminium (titanium) will have enough sensitivity to test a significant region of the parameter space of the type-I seesaw mechanism with three quasi-degenerate-in-mass heavy Majorana neutrinos~\cite{Urquia-Calderon:2022ufc}. 
 The key phenomenological parameters in cLFV searches of quasi-degenerate-in-mass HNLs are the combinations $   \Theta_{l l'}^2 \equiv \Big|\sum_{j=1}^3\Theta_{l j}^* \Theta_{l' j}\Big|$,  $l,\,l' = e,\,\mu,\,\tau$~\cite{Ibarra:2011xn,Dinh:2012bp,Alonso:2012ji}. The rates of the $\mu \rightarrow e\gamma$ and $\mu \rightarrow eee$ decays and
 $\mu$\,--\,$e$ conversion in nuclei depend on the quantity $\Theta_{\mu e}^2$. The best sensitivities come from $\mu$\,--\,$e$ conversion in titanium~\cite{Barlow:2011zza}, but the sensitivity to $\Theta^2_{\mu e}$ reaches at most $\sim$ few $\times\,10^{-8}$ for $\Mav = 100\,\text{GeV}$ (see, e.g.,~\cite{Granelli:2022eru}). Unfortunately, the mixing predicted by the models considered here are too small in magnitude to be tested at these classes of experiments, unless the precision is further enhanced in the future. 
    The prospective
sensitivities of the experiments looking at cLFV decays of
the tau-lepton, such as BELLE II~\cite{BELLEII, BELLEIIbook}, are significantly lower than those involving muons and thus are even less promising in this regard.

%%%%%%%%%%%%%%%%%%%%%%%%%%%%%%%%%%%%%%%%%%%%%%
\section{Summary and conclusions}
\label{sec:summary}
%%%%%%%%%%%%%%%%%%%%%%%%%%%%%%%%%%%%%%%%%%%%%%

In this work we have investigated a new approach to the 
low-scale type-I seesaw mechanism of neutrino mass generation, based on the modular invariance approach to lepton flavour. 
It provides a natural explanation of the mass splitting of the Heavy Neutral Leptons (heavy Majorana neutrinos) associated with the type-I seesaw mechanism, and leads to testable HNL phenomenology. 
The modular flavour models in which it is embedded 
allow also to explain the charged-lepton mass hierarchies without 
fine-tuning. Thus, it provides an unified explanation for 
the smallness of neutrino masses,  the peculiar form of 
neutrino mixing, HNL mass splitting and charged-lepton mass hierarchies.

The considered mechanism may be called a \emph{modular-symmetry-protected low-scale seesaw}. The general idea is to mimic (non-standard lepton number) L symmetry through modular symmetry, namely via an approximate residual modular symmetry.
The small L-breaking is parameterized by the \virg{distance} $\epsilon$ of the modulus from a symmetric (or fixed) point in its fundamental domain.
As a result, for the first time in the literature, the same $\epsilon$ which governs the charged-lepton mass hierarchies also participates in the suppression of light neutrino masses and in the splitting of a pseudo-Dirac HNL pair. The modulus VEV is then simultaneously responsible for the spontaneous breaking of flavour, CP and lepton number symmetries.

In~\cref{sec:modsymprot}, we establish a set of criteria that guarantee the  reliability of the mechanism. This procedure allowed us to discard a plethora of models and some modular groups. For example, under our requirements, the groups $\Gamma_2'\simeq S_3$ and $\Gamma_5'\simeq A_5'$ are excluded. The surviving cases, with the groups $\Gamma_3'\simeq T'$ and $\Gamma_4'\simeq S_4'$, are listed in~\cref{app:models} for relatively low weights, but a successful fit of the relevant low-energy data is not guaranteed. On the other hand, a good-quality description of the data was obtained for four benchmark models, discussed
in~\cref{sec:models}, where the gCP symmetry is imposed and thus the modulus is the only source of CP violation. These models can involve as little as 6 dimensionless superpotential constant parameters.

Due to the proximity of $\tau$ to the CP-conserving symmetric points, the CP violation in the lepton sector turns out to be quite small in the considered benchmark models. This feature will be 
tested
in forthcoming long-baseline and reactor neutrino oscillation experiments. 
However, if one gives up on imposing the gCP symmetry from the start (and thus gives up on minimality), a fit with a relatively large Dirac phase can be obtained.
We have discussed in detail the HNL phenomenology in~\cref{sec:pheno}.
We find that not only do the considered models predict HNL mixings within the reach of future experiments, but that they also give rise to distinct flavour structures, singling out nearly point-like regions within the flavour triangles. These features allow to test the models and distinguish between them.

The proposed link between modular symmetry and L symmetry, in the context of the considered low-scale type-I seesaw mechanism, which connects charged-lepton masses, neutrino masses, neutrino mixing, leptonic CP violation and HNL phenomenology, while also avoiding fine-tuning and retaining minimality and predictivity, opens up the possibility to probe the modular flavour paradigm in new ways.

\vskip 5mm

%%%%%%%%%%%%%%%%%%%%%%%%%%%%%%%%%%%%%%%%%%%%%%
\section*{Acknowledgements}
%%%%%%%%%%%%%%%%%%%%%%%%%%%%%%%%%%%%%%%%%%%%%%

This work was supported in part by the European Union's Horizon research and innovation programme under the Marie Sk\l{}odowska-Curie grant agreements No.~860881-HIDDeN and No.~101086085-ASYMMETRY, and by the INFN program on Theoretical Astroparticle Physics. The work of S.T.P.~was supported also by the World Premier International Research Center Initiative (WPI Initiative, MEXT), Japan.

\vfill
\clearpage

\appendix

%%%%%%%%%%%%%%%%%%%%%%%%%%%%%%%%%%%%%%%%%%%%%%
\section{Normalized modular forms} 
\label{app:forms}
%%%%%%%%%%%%%%%%%%%%%%%%%%%%%%%%%%%%%%%%%%%%%%

Here we present, for completeness, the $q$-expansions up to $\mathcal{O}(q^{4})$ of the modular form multiplets used in the text ($q = e^{2\pi i \tau}$), normalized according to the Petersson prescription in Ref.~\cite{Petcov:2023fwh}. We have checked that, for scans inside the fundamental domain, higher-order terms have a negligible effect.
For $\Gamma_3^{(\prime)} \simeq A_4^{(\prime)}$, the relevant modular form triplets are:
\begin{equation} \label{eq:A4triplets}
\begin{aligned}
        Y^{(2)}_{\mathbf{3}} &\simeq
0.564\left(
\begin{array}{c}
 1+12 \,q+36 \,q^2+12 \,q^3+84 \,q^4\\
 -6 \,q^{1/3}-42 \,q^{4/3}-48 \,q^{7/3}-108 \,q^{10/3}\\
 -18 \,q^{2/3}-36 \,q^{5/3}-90 \,q^{8/3}-72 \,q^{11/3} \\
\end{array}
\right)\,,
        \\[2mm]
        Y^{(4)}_{\mathbf{3}} &\simeq
        0.846\left(
\begin{array}{c}
 1-84 \,q-756 \,q^2-2028 \,q^3-6132 \,q^4 \\
 6 \,q^{1/3}+438 \,q^{4/3}+2064 \,q^{7/3}+6804 \,q^{10/3}\\
 54 \,q^{2/3}+756 \,q^{5/3}+3510 \,q^{8/3}+7992 \,q^{11/3} \\
\end{array}
\right)\,,
        \\[2mm]
        Y^{(6)}_{\mathbf{3},1} &\simeq
        1.058\left(
\begin{array}{c}
 18 \,q-108 \,q^2+162 \,q^3+72 \,q^4\\
 -\,q^{1/3}-4 \,q^{4/3}+40 \,q^{7/3}+36 \,q^{10/3}\\
 6 \,q^{2/3}-6 \,q^{5/3}-168 \,q^{8/3}+564 \,q^{11/3}\\
\end{array}
\right)\,,
        \\[2mm] 
        Y^{(6)}_{\mathbf{3},2} &\simeq
        0.144\left(
\begin{array}{c}
 1+36 \,q+6372 \,q^2+39348 \,q^3+178020 \,q^4 \\
 6 \,q^{1/3}-1434 \,q^{4/3}-23568 \,q^{7/3}-143100 \,q^{10/3}\\
 -90 \,q^{2/3}-4284 \,q^{5/3}-45594 \,q^{8/3}-227160 \,q^{11/3}\\
\end{array}
\right)\,.
\end{aligned}
\end{equation}
For $\Gamma_4' \simeq S_4'$, instead, we have:
\begin{equation}
        Y^{(2)}_{\mathbf{2}} \simeq 
        0.532 \left(
\begin{array}{c}
 1+24 \,q+24 \,q^2+96 \,q^3+24 \,q^4 \\
 -8 \sqrt{3} \,q^{1/2}-32 \sqrt{3} \,q^{3/2}-48 \sqrt{3} \,q^{5/2}-64 \sqrt{3} \,q^{7/2}\\
\end{array}
\right)\,,
\end{equation}
for the relevant doublet modular form, and
\begin{equation} \label{eq:normalizedtriplets}
\begin{aligned}
        Y^{(3)}_{\mathbf{\hat{3}}} &\simeq
0.580 \left(
\begin{array}{c}
 4 \sqrt{2}\, q^{1/4}+104 \sqrt{2}\, q^{5/4}+292 \sqrt{2}\, q^{9/4}+680 \sqrt{2}\, q^{13/4} \\
 20 \,q^{1/2}+96\, q^{3/2}+520\, q^{5/2}+576\, q^{7/2}\\
 1-68\, q-260\, q^2-480\, q^3-1028\, q^4\\
\end{array}
\right)\,,
        \\[2mm]
        Y^{(3)}_{\mathbf{\hat{3}'}} &\simeq
0.580 \left(
\begin{array}{c}
 -32 \sqrt{2}\, q^{3/4}-192 \sqrt{2}\, q^{7/4}-480 \sqrt{2}\, q^{11/4}-832 \sqrt{2}\, q^{15/4} \\
 1+60\, q+252\, q^2+544\, q^3+1020\, q^4\\
 -12 \,q^{1/2}-160\, q^{3/2}-312\, q^{5/2}-960\, q^{7/2} \\
\end{array}
\right)\,,
        \\[2mm]
        Y^{(4)}_{\mathbf{3}} &\simeq
        1.407 \left(
\begin{array}{c}
 2 \sqrt{2}\, q^{1/2}-8 \sqrt{2} \,q^{3/2}-4 \sqrt{2} \,q^{5/2}+48 \sqrt{2} \,q^{7/2} \\
 4 \,q^{3/4}-24 \,q^{7/4}+44 \,q^{11/4}-8 \,q^{15/4} \\
 -q^{1/4}+2 \,q^{5/4}+11 \,q^{9/4}-22 \,q^{13/4}\\
\end{array}
\right)\,,
        \\[2mm]
        Y^{(5)}_{\mathbf{\hat{3}},1} &\simeq
1.084 \left(
\begin{array}{c}
 -512 \sqrt{2} \,q^{5/4}-5120 \sqrt{2} \,q^{9/4}-23040 \sqrt{2} \,q^{13/4} \\
 -20 \,q^{1/2}-960 \,q^{3/2}-8424 \,q^{5/2}-28800 \,q^{7/2}\\
 1+180 \,q+3380 \,q^2+16320 \,q^3+52020 \,q^4 \\
\end{array}
\right)\,, 
        \\[2mm]
        Y^{(5)}_{\mathbf{\hat{3}},2} &\simeq
0.729 \left(
\begin{array}{c}
 -q^{1/4}+14 \,q^{5/4}-81 \,q^{9/4}+238 \,q^{13/4}\\
 4 \sqrt{2} \,q^{1/2}-56 \sqrt{2} \,q^{5/2}\\
 16 \sqrt{2} \,q-64 \sqrt{2} \,q^2+256 \sqrt{2} \,q^4\\
\end{array}
\right)\,,
        \\[2mm]
        Y^{(5)}_{\mathbf{\hat{3}'}} &\simeq
0.801 \left(
\begin{array}{c}
 64 \sqrt{2}\, q^{3/4}+1920 \sqrt{2}\, q^{7/4}+11712 \sqrt{2}\, q^{11/4}+40064 \sqrt{2}\, q^{15/4}\\
 1-204\, q-3276\, q^2-16448\, q^3-52428\, q^4 \\
 12\, q^{1/2}+1088\, q^{3/2}+7512\, q^{5/2}+32640\, q^{7/2} \\
\end{array}
\right) \,,
\end{aligned}
\end{equation}
for the triplet modular forms.

\vfill
\clearpage

%%%%%%%%%%%%%%%%%%%%%%%%%%%%%%%%%%%%%%%%%%%%%%
\section{Model landscape} 
\label{app:models}
%%%%%%%%%%%%%%%%%%%%%%%%%%%%%%%%%%%%%%%%%%%%%%

%%%%%%%%%%%%%%%%%%%%%%
\begin{table}[h]
\centering
\renewcommand{\arraystretch}{1.2}
\caption{
Inequivalent $T'$-based symmetry-protected seesaw models passing the criteria of~\cref{sec:modsymprot}, for $\tau \simeq \tau_\text{sym}$. The $\Delta M$ column indicates the expected magnitude of the pseudo-Dirac HNL splitting. We impose the upper bound $k_Y \leq 6$ on modular form weights.
}
\label{tab:modelsA4}
\begin{tabular}{x{2cm}x{6cm}x{1.5cm}x{1.5cm}x{2cm}} 
\toprule
$\tau_\text{sym}$&
$L$ & $E^c$ & $N^c$ & $\Delta M$
 \\ 
\midrule
\multirow{11}{*}{$\omega$} &
$(\mathbf{1''}, 0) \oplus(\mathbf{1}, +2) \oplus(\mathbf{1'}, +4)$ & $(\mathbf{3},+2)$ & $(\mathbf{\hat{2}},+1)$ & \multirow{3}{*}{$\sim \epsilon\, M$} \\
& $(\mathbf{1}, -1) \oplus(\mathbf{1'}, +1) \oplus(\mathbf{1''}, +3)$ & $(\mathbf{3},+3)$ & $(\mathbf{\hat{2}''},+2)$ & \\
& $(\mathbf{1'}, -2) \oplus(\mathbf{1''}, 0) \oplus(\mathbf{1}, +2)$ & $(\mathbf{3},+4)$ & $(\mathbf{\hat{2}'},+3)$  & \\
  \cmidrule{2-5}
& $(\mathbf{1''}, +2) \oplus(\mathbf{1}, +4) \oplus(\mathbf{1'}, +6)$ & \multirow{2}{*}{$(\mathbf{3},0)$} & \multirow{2}{*}{$(\mathbf{3},0)$} & \multirow{2}{*}{$\sim m_\nu$} \\
& $(\mathbf{1}, +2) \oplus(\mathbf{1'}, +4) \oplus(\mathbf{1''}, +6)$ & & & \\
  \cmidrule{2-5}
& $(\mathbf{1}, +1) \oplus(\mathbf{1'}, +3) \oplus(\mathbf{1''}, +5)$ & \multirow{2}{*}{$(\mathbf{3},+1)$} & \multirow{2}{*}{$(\mathbf{3},+1)$} & \multirow{6}{*}{$\sim \epsilon\, M$} \\
& $(\mathbf{1'}, +1) \oplus(\mathbf{1''}, +3) \oplus(\mathbf{1}, +5)$ & & & \\
& $(\mathbf{1'}, 0) \oplus(\mathbf{1''}, +2) \oplus(\mathbf{1}, +4)$ & \multirow{2}{*}{$(\mathbf{3},+2)$} & \multirow{2}{*}{$(\mathbf{3},+2)$} & \\
& $(\mathbf{1''}, 0) \oplus(\mathbf{1}, +2) \oplus(\mathbf{1'}, +4)$ & & & \\
& $(\mathbf{1''}, -1) \oplus(\mathbf{1}, +1) \oplus(\mathbf{1'}, +3)$ & \multirow{2}{*}{$(\mathbf{3},+3)$} & \multirow{2}{*}{$(\mathbf{3},+3)$} & \\
& $(\mathbf{1}, -1) \oplus(\mathbf{1'}, +1) \oplus(\mathbf{1''}, +3)$ & && \\
\midrule
\multirow{11}{*}{ $i\infty$} &
$(\mathbf{1'}, 0) \oplus(\mathbf{1'}, +2) \oplus(\mathbf{1'}, +4)$ & $(\mathbf{3},+2)$ & $(\mathbf{\hat{2}'},+1)$ & \multirow{3}{*}{$\sim \epsilon\, M$} \\ &
$(\mathbf{1'}, -1) \oplus(\mathbf{1'}, +1) \oplus(\mathbf{1'}, +3)$ & $(\mathbf{3},+3)$ & $(\mathbf{\hat{2}'},+2)$ & \\ &
$(\mathbf{1'}, -2) \oplus(\mathbf{1'}, 0) \oplus(\mathbf{1'}, +2)$ & $(\mathbf{3},+4)$ & $(\mathbf{\hat{2}'},+3)$ & \\ 
  \cmidrule{2-5}
&
$(\mathbf{1'}, +2) \oplus(\mathbf{1'}, +4) \oplus(\mathbf{1'}, +6)$ & \multirow{2}{*}{$(\mathbf{3},0)$} & \multirow{2}{*}{$(\mathbf{3},0)$} & \multirow{2}{*}{$\sim m_\nu$} \\ 
& $(\mathbf{1''}, +2) \oplus(\mathbf{1''}, +4) \oplus(\mathbf{1''}, +6)$ & && \\ 
  \cmidrule{2-5}
& $(\mathbf{1'}, +1) \oplus(\mathbf{1'}, +3) \oplus(\mathbf{1'}, +5)$ & \multirow{2}{*}{$(\mathbf{3},+1)$} & \multirow{2}{*}{$(\mathbf{3},+1)$} & \multirow{6}{*}{$\sim \epsilon\, M$}\\ &
$(\mathbf{1''}, +1) \oplus(\mathbf{1''}, +3) \oplus(\mathbf{1''}, +5)$ & & & \\ &
$(\mathbf{1'}, 0) \oplus(\mathbf{1'}, +2) \oplus(\mathbf{1'}, +4)$ & \multirow{2}{*}{$(\mathbf{3},+2)$} & \multirow{2}{*}{$(\mathbf{3},+2)$} & \\ &
$(\mathbf{1''}, 0) \oplus(\mathbf{1''}, +2) \oplus(\mathbf{1''}, +4)$ & && \\ &
$(\mathbf{1'}, -1) \oplus(\mathbf{1'}, +1) \oplus(\mathbf{1'}, +3)$ & \multirow{2}{*}{$(\mathbf{3},+3)$} & \multirow{2}{*}{$(\mathbf{3},+3)$} & \\ &
$(\mathbf{1''}, -1) \oplus(\mathbf{1''}, +1) \oplus(\mathbf{1''}, +3)$ & & &
  \\
\bottomrule
\end{tabular}
\end{table}
\renewcommand{\arraystretch}{1.0}
%%%%%%%%%%%%%%%%%%%%%%
%

\vfill
\clearpage

%%%%%%%%%%%%%%%%%%%%%%
\begin{table}[p]
\centering
\renewcommand{\arraystretch}{1.2}
\caption{
Inequivalent $S_4'$-based symmetry-protected seesaw models passing the criteria of~\cref{sec:modsymprot}, for $\tau \simeq i \infty$. The $\Delta M$ column indicates the expected magnitude of the pseudo-Dirac HNL splitting. We impose the upper bound $k_Y \leq 5$ on modular form weights.
}
\label{tab:modelsS4}
\begin{tabular}{x{5cm}x{5cm}x{1.5cm}x{2cm}} 
\toprule
$L$ & $E^c$ & $N^c$ & $\Delta M$
 \\ 
\midrule
\multirow{2}{*}{$(\mathbf{\hat{3}},+1)$} & $(\mathbf{\hat{1}'}, +3) \oplus(\mathbf{\hat{2}}, +3)$ & \multirow{8}{*}{$(\mathbf{\hat{2}},+1)$} & \multirow{16}{*}{$\sim \epsilon^2\, M$} \\
 & $(\mathbf{1}, +2) \oplus(\mathbf{\hat{1}'}, +3) \oplus(\mathbf{1}, +4)$ &  & \\
\multirow{2}{*}{$(\mathbf{\hat{3}'},+1)$} & $(\mathbf{\hat{1}}, +3) \oplus(\mathbf{\hat{2}}, +3)$ &  & \\
 & $(\mathbf{1}, 0) \oplus(\mathbf{1}, +2) \oplus(\mathbf{1}, +4)$ &  & \\
\multirow{2}{*}{$(\mathbf{\hat{3}},+3)$} & $(\mathbf{\hat{1}'}, +1) \oplus(\mathbf{\hat{2}}, +1)$ &  & \\
 & $(\mathbf{1}, 0) \oplus(\mathbf{\hat{1}'}, +1) \oplus(\mathbf{1}, +2)$ &  & \\
\multirow{2}{*}{$(\mathbf{\hat{3}'},+3)$} & $(\mathbf{\hat{1}}, +1) \oplus(\mathbf{\hat{2}}, +1)$ &  & \\
 & $(\mathbf{1}, -2) \oplus(\mathbf{1}, 0) \oplus(\mathbf{1}, +2)$ &  & \\
\cmidrule{1-3}
\multirow{2}{*}{$(\mathbf{\hat{3}},0)$} & $(\mathbf{\hat{1}'}, +4) \oplus(\mathbf{\hat{2}}, +4)$ & \multirow{8}{*}{$(\mathbf{\hat{2}},+2)$} & \\
 & $(\mathbf{1}, +3) \oplus(\mathbf{\hat{1}'}, +4) \oplus(\mathbf{1}, +5)$ &  & \\
\multirow{2}{*}{$(\mathbf{\hat{3}'},0)$} & $(\mathbf{\hat{1}}, +4) \oplus(\mathbf{\hat{2}}, +4)$ &  & \\
 & $(\mathbf{1}, +1) \oplus(\mathbf{1}, +3) \oplus(\mathbf{1}, +5)$ &  & \\
\multirow{2}{*}{$(\mathbf{\hat{3}},+2)$} & $(\mathbf{\hat{1}'}, +2) \oplus(\mathbf{\hat{2}}, +2)$ &  & \\
 & $(\mathbf{1}, +1) \oplus(\mathbf{\hat{1}'}, +2) \oplus(\mathbf{1}, +3)$ &  & \\
\multirow{2}{*}{$(\mathbf{\hat{3}'},+2)$} & $(\mathbf{\hat{1}}, +2) \oplus(\mathbf{\hat{2}}, +2)$ &  & \\
 & $(\mathbf{1}, -1) \oplus(\mathbf{1}, +1) \oplus(\mathbf{1}, +3)$ &  & \\
\midrule
$(\mathbf{\hat{1}}, +3) \oplus(\mathbf{1}, +4) \oplus(\mathbf{\hat{1}}, +5)$ & \multirow{2}{*}{$(\mathbf{3},0)$} & \multirow{2}{*}{$(\mathbf{3},0)$} & \multirow{3}{*}{$\sim m_\nu$} \\
$(\mathbf{\hat{1}'}, +3) \oplus(\mathbf{1}, +4) \oplus(\mathbf{\hat{1}'}, +5)$ & & & \\
$(\mathbf{\hat{1}}, +3) \oplus(\mathbf{1'}, +4) \oplus(\mathbf{\hat{1}}, +5)$ & $(\mathbf{3'},0)$ & $(\mathbf{3'},0)$ & \\
\midrule
$(\mathbf{\hat{1}}, +2) \oplus(\mathbf{1}, +3) \oplus(\mathbf{\hat{1}}, +4)$ & \multirow{2}{*}{$(\mathbf{3},+1)$} & \multirow{2}{*}{$(\mathbf{3},+1)$} & \multirow{6}{*}{$\sim \epsilon^2\, M$} \\
$(\mathbf{\hat{1}'}, +2) \oplus(\mathbf{1}, +3) \oplus(\mathbf{\hat{1}'}, +4)$ & & & \\
$(\mathbf{\hat{1}}, +2) \oplus(\mathbf{1'}, +3) \oplus(\mathbf{\hat{1}}, +4)$ & $(\mathbf{3'},+1)$ & $(\mathbf{3'},+1)$ & \\
$(\mathbf{\hat{1}}, +1) \oplus(\mathbf{1}, +2) \oplus(\mathbf{\hat{1}}, +3)$ & \multirow{2}{*}{$(\mathbf{3},+2)$} & \multirow{2}{*}{$(\mathbf{3},+2)$} & \\
$(\mathbf{\hat{1}'}, +1) \oplus(\mathbf{1}, +2) \oplus(\mathbf{\hat{1}'}, +3)$ & & & \\
$(\mathbf{\hat{1}}, +1) \oplus(\mathbf{1'}, +2) \oplus(\mathbf{\hat{1}}, +3)$ & $(\mathbf{3'},+2)$ & $(\mathbf{3'},+2)$ \\
\bottomrule
\end{tabular}
\end{table}
\renewcommand{\arraystretch}{1.0}
%%%%%%%%%%%%%%%%%%%%%%
%
\vfill
\clearpage

%%%%%%%%%%%%%%%%%%%%%%%%%%%%%%%%%%%%%%%%%%%%%%
\section{Analytical formulae for mass ratios} 
\label{app:analytics}
%%%%%%%%%%%%%%%%%%%%%%%%%%%%%%%%%%%%%%%%%%%%%%

\paragraph{Models A and B.}

The charged-lepton mass matrix in~\cref{eq:YeA} is common to both these models and it is written in the $ST$-diagonal basis. Hence, to obtain analytic expressions for the charged-lepton masses, at leading order in $|\epsilon|$, where $\epsilon = (\tau-\omega)/(\tau-\omega^2)$, one must take into account the change-of-basis matrix of~\cref{foot:ST} and the $q$-expansions of~\cref{app:forms}. We find:
\begin{equation} \label{eq:mlAB}
\begin{aligned}
    m_\tau &\,\simeq\,\sqrt{\tilde\alpha_1^2 + \tilde\alpha_2^2 + (\tilde\alpha_{3,1}-\tilde\alpha_{3,2})^2}\, v_d \,,\\
    m_\mu &\,\simeq\,4\sqrt{\frac{5}{3}}\,|\tilde\alpha_{3,2}|\,\sqrt{\frac{\tilde\alpha_1^2 + \tilde\alpha_2^2}{\tilde\alpha_1^2 + \tilde\alpha_2^2 + (\tilde\alpha_{3,1}-\tilde\alpha_{3,2})^2}}\, |\epsilon|\, v_d \,,\\
    m_e &\,\simeq\,  \frac{40}{3}\,
    \frac{\tilde\alpha_1|\tilde\alpha_2|}{\sqrt{\tilde\alpha_1^2 + \tilde\alpha_2^2}}\,|\epsilon|^2\, v_d \,,
\end{aligned}
\end{equation}
in agreement with the expected spectrum hierarchy.
Here, $\tilde\alpha_i \equiv \mathcal{N}_i \,\alpha_i$, where 
$\mathcal{N}_1 \simeq 0.803$, $\mathcal{N}_2 \simeq 1.713$, $\mathcal{N}_{3,1} \simeq 0.254$ and $\mathcal{N}_{3,2} \simeq 0.415$.
Recall that the constants $\alpha_i$ already include the effects of canonical field normalization, as discussed at the start of~\cref{sec:modsymprot}.

\paragraph{Model C.}

The charged-lepton mass matrix in~\cref{eq:YeC} is written in the $T$-diagonal basis. One can directly use the $q$-expansions given in~\cref{app:forms} to obtain analytic expressions for the charged-lepton masses, at leading order in $|\epsilon| = e^{-(\pi/ 2 )\im\tau }$, with $\epsilon = q^{1/4}$.
We find:
\begin{equation} \label{eq:mlC}
\begin{aligned}
    m_\tau &\,\simeq\,\sqrt{\tilde\alpha_1^2 + \tilde\alpha_3^2}\, v_d \,,\\
    m_\mu &\,\simeq\,|\tilde\alpha_2|\, |\epsilon|\, v_d \,,\\
    m_e &\,\simeq\,  144\sqrt{2}\,
    \frac{\tilde\alpha_1|\tilde\alpha_3|}{\sqrt{\tilde\alpha_1^2 + \tilde\alpha_3^2}}|\epsilon|^3\, v_d \,,
\end{aligned}
\end{equation}
in agreement with the expected spectrum hierarchy.
Here, $\tilde\alpha_i \equiv \mathcal{N}_i \,\alpha_i$, where 
$\mathcal{N}_1 \simeq 0.580$, $\mathcal{N}_2 \simeq 1.407$ and $\mathcal{N}_{3} \simeq 0.801$ are the overall normalization factors of 
$Y^{(3)}_{\mathbf{\hat{3}'}}$, $Y^{(4)}_{\mathbf{3}}$ and 
$Y^{(5)}_{\mathbf{\hat{3}'}}$, respectively, cf.~\cref{eq:normalizedtriplets}.
The $\alpha_i$ include the effects of canonical normalization.

\paragraph{Model D.}

The charged-lepton mass matrix in~\cref{eq:YeD} is also written in the $T$-diagonal basis. One can once again directly use the $q$-expansions of~\cref{app:forms} to obtain analytic expressions for the charged-lepton masses, at leading order in $|\epsilon| = e^{-(\pi/ 2 )\im\tau }$.
We find:
\begin{equation} \label{eq:mlD}
\begin{aligned}
    m_\tau &\,\simeq\,\sqrt{\tilde\alpha_1^2 + \tilde\alpha_{3,1}^2}\, v_d \,,\\
    m_\mu &\,\simeq\,\tilde\alpha_1\frac{|4\sqrt{2}\,\tilde\alpha_{3,1}+\tilde\alpha_{3,2}|}{\sqrt{\tilde\alpha_1^2 + \tilde\alpha_{3,1}^2}} |\epsilon|\, v_d \,,\\
    m_e &\,\simeq\,  72\,|\tilde\alpha_{2}|\left|\frac{2\sqrt{2}\,\tilde\alpha_{3,1}-\tilde\alpha_{3,2}}{4\sqrt{2}\,\tilde\alpha_{3,1}+\tilde\alpha_{3,2}}\right| |\epsilon|^3\, v_d \,,
\end{aligned}
\end{equation}
in agreement with the expected spectrum hierarchy.
Here, $\tilde\alpha_i \equiv \mathcal{N}_i \,\alpha_i$, where 
$\mathcal{N}_1 \simeq 0.580$, $\mathcal{N}_2 \simeq 1.407$, $\mathcal{N}_{3,1} \simeq 1.084$ and $\mathcal{N}_{3,2} \simeq 0.729$ are the overall normalization factors of 
$Y^{(3)}_{\mathbf{\hat{3}}}$, $Y^{(4)}_{\mathbf{3}}$, 
$Y^{(5)}_{\mathbf{\hat{3}},1}$ and 
$Y^{(5)}_{\mathbf{\hat{3}},2}$, cf.~\cref{eq:normalizedtriplets}. 
The $\alpha_i$ include the effects of canonical normalization.

\vfill
\clearpage

\footnotesize
%%%%%%%%%%%%%%%%%%%%%%%
\bibliographystyle{JHEPwithnote}
\bibliography{bibliography}
%%%%%%%%%%%%%%%%%%%%%%%

\end{document}